\documentclass[fleqn,10pt]{wlscirep}
\usepackage[utf8]{inputenc}
\usepackage[T1]{fontenc}
\usepackage{graphicx}
\usepackage{epstopdf}
\usepackage{subfigure}
\title{Understanding urban congestion with biking traffic and routing detour ratio}

\author[a]{Xinze Qiu}
\author[a]{Tianli Gao}
\author[a]{Yu Yang}
\author[a]{Ankang Luo}
\author[a,*]{Fan Shang}
\author[a,*]{Ruiqi Li}

\affil[a]{UrbanNet Lab, College of Information Science and Technology, Beijing University of Chemical Technology, Beijing 100029, China}

\affil[*]{corresponding authors. \textit{E-mail address:} lir@buct.edu.cn (R. Li), sf@buct.edu.cn (F. Shang).}

\begin{abstract}
Bike-sharing systems have been regarded as a critical component of solutions towards the transition to a greener and more sustainable transportation, with the benefits of reducing carbon emissions, improving public health, and mitigating congestion by replacing short-distance motorized trips. 
Due to better accessibility and usage flexibility, newly emergent dockless sharing bikes have become quite popular and are reviving the fashion of cycling in cities. 
Urban congestion is simultaneously influenced by heterogeneous saptio-temporal travel demands, topology and spatial characteristics of road networks, and the interplay between travel modes. 
In this paper, by considering aforementioned factors, we discover a robust sublinear scaling relation between the level of congestion for vehicles and detour ratio weighted by biking traffic, which is intriguing given the fact that congestion and detour ratio are linearly independent. Such a scaling relation implies a strong interplay between vehicle traffic and cycling activities, and can be applied in predictions for congestion or aggregated to more sophisticated traffic models. 
In addition, biking-traffic-weighted detour ratio can be applied to detect inefficient routes, 
which would help alleviate urban congestion, make better urban planning, and improve transportation efficiency and equity in cities. 
\end{abstract}


\begin{document}

\flushbottom
\maketitle

\section{Introduction}

Most modern cities have followed a car-centric development since the 20th-century \cite{jacobs1961death}, 
and the booming of cars and car-orientated transportation systems lead to a variety of urban illnesses, including traffic congestion, air pollution, energy deficiency, and deterioration of public health \cite{WHO2002physically, demaio2009bike,jappinen2013modelling}. 
Though with an ideal of providing a solution to sustainable mobility, 
car-sharing platforms have intensified traffic congestion on both intensity and duration since their launching \cite{diao2021impacts}. 
Over the past few decades, with growing concerns over global warming and rapid urban sprawl \cite{li2017simple}, numerous efforts have been devoted to promoting public bike-sharing systems in cities as a greener, more resilient, and healthier mobility solution to reduce carbon emissions and improve public health \cite{dill2003bicycle,jappinen2013modelling,hull2014bicycle,2019bikelife,cheng2021role}. 
During the COVID-19 pandemic, bike-sharing systems are proved to be more resilient \cite{teixeira2020link} and safer to move around for essential needs, since biking allows for greater social distancing than other means of public transportation \cite{jiang2020dockless,li2018effect,2020meninoSurvey}. 
With great advances in IoT (Internet of Things) and mobile payment technology, and attributed to its better affordability \cite{gao2022quantifying}, accessibility, and usage flexibility \cite{li2021gravity}, newly emergent dockless sharing bikes \cite{sun2018sharing,jiang2020dockless,li2022emergence} have become quite popular and revived the riding fashion with dozens of millions of biking trips over hundreds of cities \cite{jiang2020dockless}. 

It is commonly believed that biking has a potential of replacing short-distance motorized trips \cite{martin2014evaluating, shaheen2013public} with a myriad of benefits to both individuals and the whole city, including increasing fitness and reducing the stress of riders from cycling activities \cite{cavill2006physical, rojas2011health, shaheen2010bikesharing, shaheen2013public}, saving parking space and fossil energy \cite{gossling2016urban, szell2018crowdsourced}, and mitigating urban congestion \cite{hamilton2018bicycle}. 
Urban congestion and transportation efficiency are simultaneously influenced by the topology and spatial characteristics of road networks \cite{yang2018universal}, heterogeneous spatio-temporal travel demands \cite{dong2016population}, and the interplay between travel modes (e.g., driving and cycling). 
The detour ratio, which equals the shortest routing distance on road networks over Euclidean distance between locations, reflects how many additional detours are needed to get from one place to another \cite{dong2016population,yang2018universal}. 
The sum of detour ratios has been widely adopted as a measure of transportation efficiency \cite{cardillo2006structural,gastner2006shape}. The greater the detour ratio, the lower the efficiency of transportation process. 
For a more comprehensive evaluation, heterogeneous travel demands also need to be taken into considerations \cite{dong2016population}. Inefficient routes that are used by many people would have stronger impacts on urban congestion \cite{dong2016population}. 
In addition, different travel modes might have influence over each other. 
Previous evidence indicate that promoting public bike-sharing systems can reduce car usage \cite{demaio2009bike, martin2014evaluating}, and increase the share of public transportation via covering the last mile more easily by cycling \cite{fishman2013bike, noland2006smart, shaheen2013public}. 
However, there have been few works considering all above factors in a cohesive way for gaining a better understanding of urban congestion.

In this work, by exploiting massive cycling activities via dockless sharing bikes and detailed vehicle congestion data over three diversified cities in China, Beijing, Shanghai, and Xiamen, we find that the distributions of detour ratios on cycling in different cities are quite similar, and the detour ratio is negatively correlated with Euclidean distance over cities.
More profoundly, we discover a sublinear scaling relation between the level of congestion and detour ratio weighted by biking traffic, which is intriguing given the fact that the level of congestion and detour ratio are almost linearly independent. 
In practice, biking traffic and vehicle traffic would highly probably influence each other especially when there are no dedicated biking lanes or when road conditions are complex. Such a relation implies a strong interplay between vehicle traffic and cycling activities and topology of road networks, and can be applied in predictions for congestion or aggregated to more sophisticated traffic models. 
In addition, biking-traffic-weighted detour ratio can be applied to detect inefficient routes. 




The rest of the paper is organized as follows. Section 2.1 presents some basic analysis of characteristics of biking behaviors via dockless sharing bikes.
Section 2.2 shows the distribution of detour ratios for biking trips during rush hours and off-peak hours over different cities, and explores the relationship between the detour ratio and Euclidean distance.
Section 2.3 reveals the sublinear relation between the congestion level and detour ratio weighted by biking traffic, and analyzes possible relationships between biking flow, vehicle traffic, and road structure to reveal underlying mechanics. 
Section 2.4 identifies inefficient routes via the biking-traffic-weighted DR. 
Section 3 concludes the work, points out limitations, and highlights future research directions. 

\begin{figure}[htbp] \centering
\subfigure[Beijing]
{    \begin{minipage}[b]{.32\linewidth}         \centering
        \includegraphics[scale=0.25]{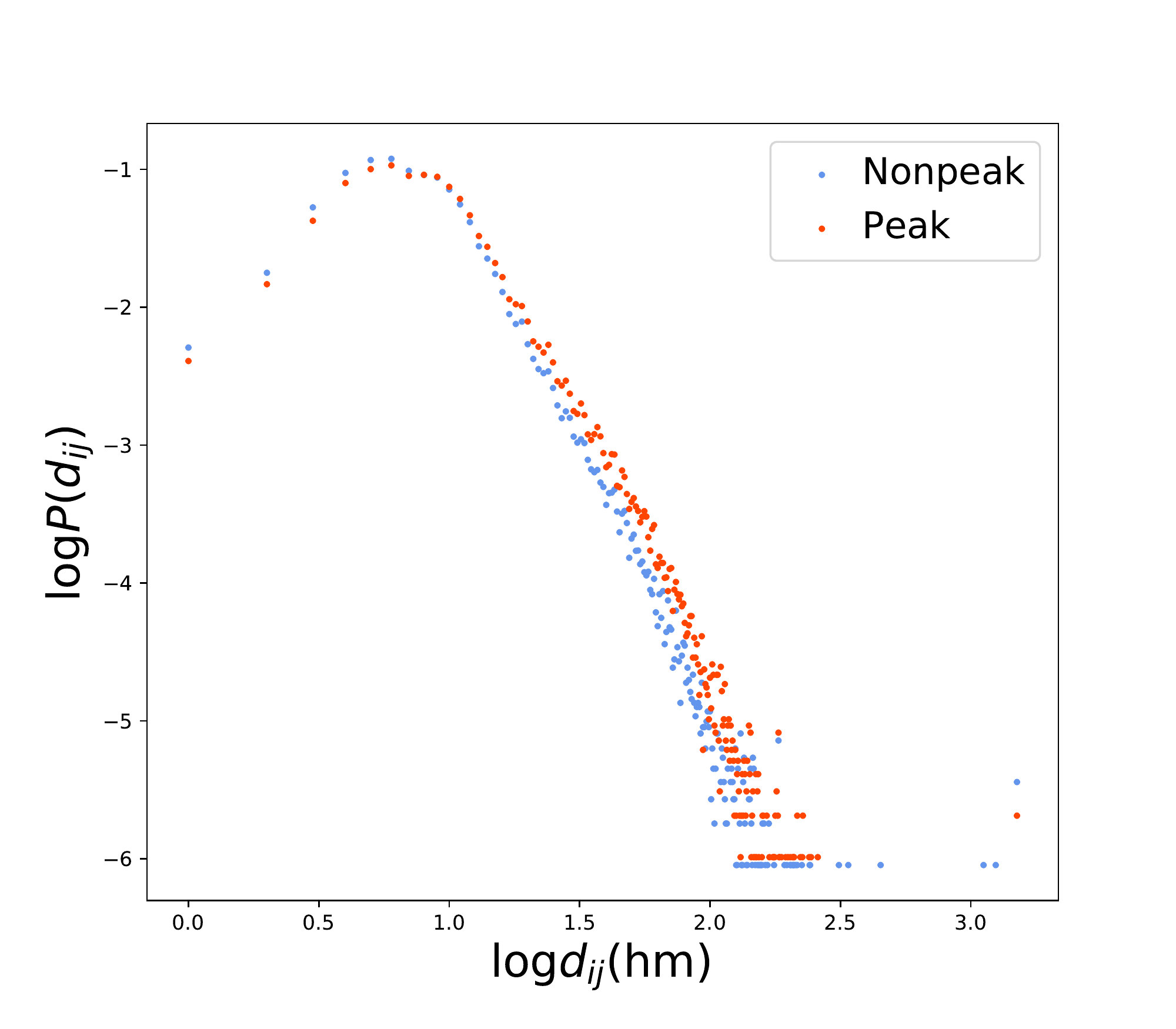}
    \end{minipage}}
\subfigure[Shanghai]
{ 	\begin{minipage}[b]{.32\linewidth}        \centering
        \includegraphics[scale=0.25]{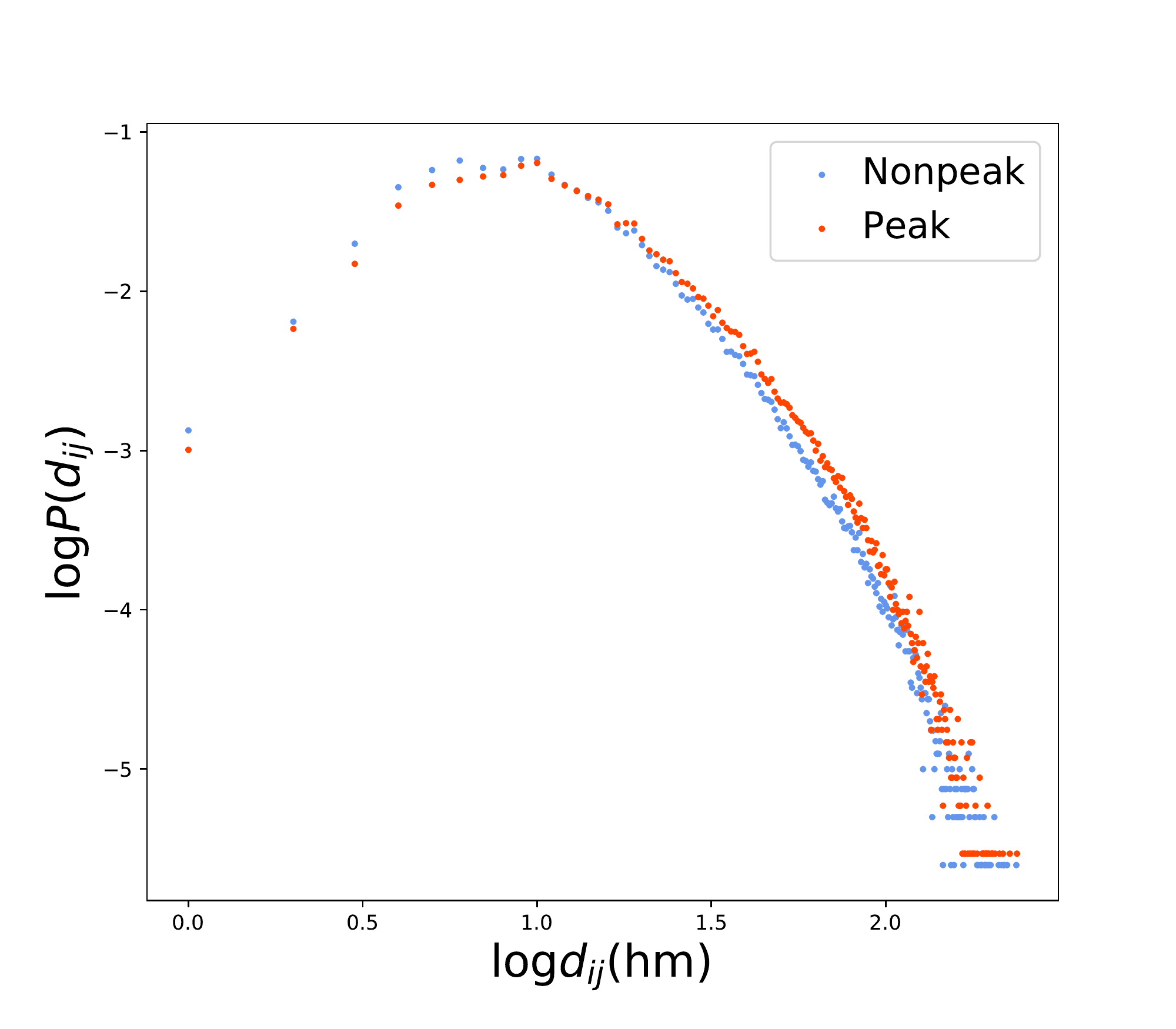}
    \end{minipage}}
\subfigure[Xiamen]
{ 	\begin{minipage}[b]{.32\linewidth}        \centering
        \includegraphics[scale=0.25]{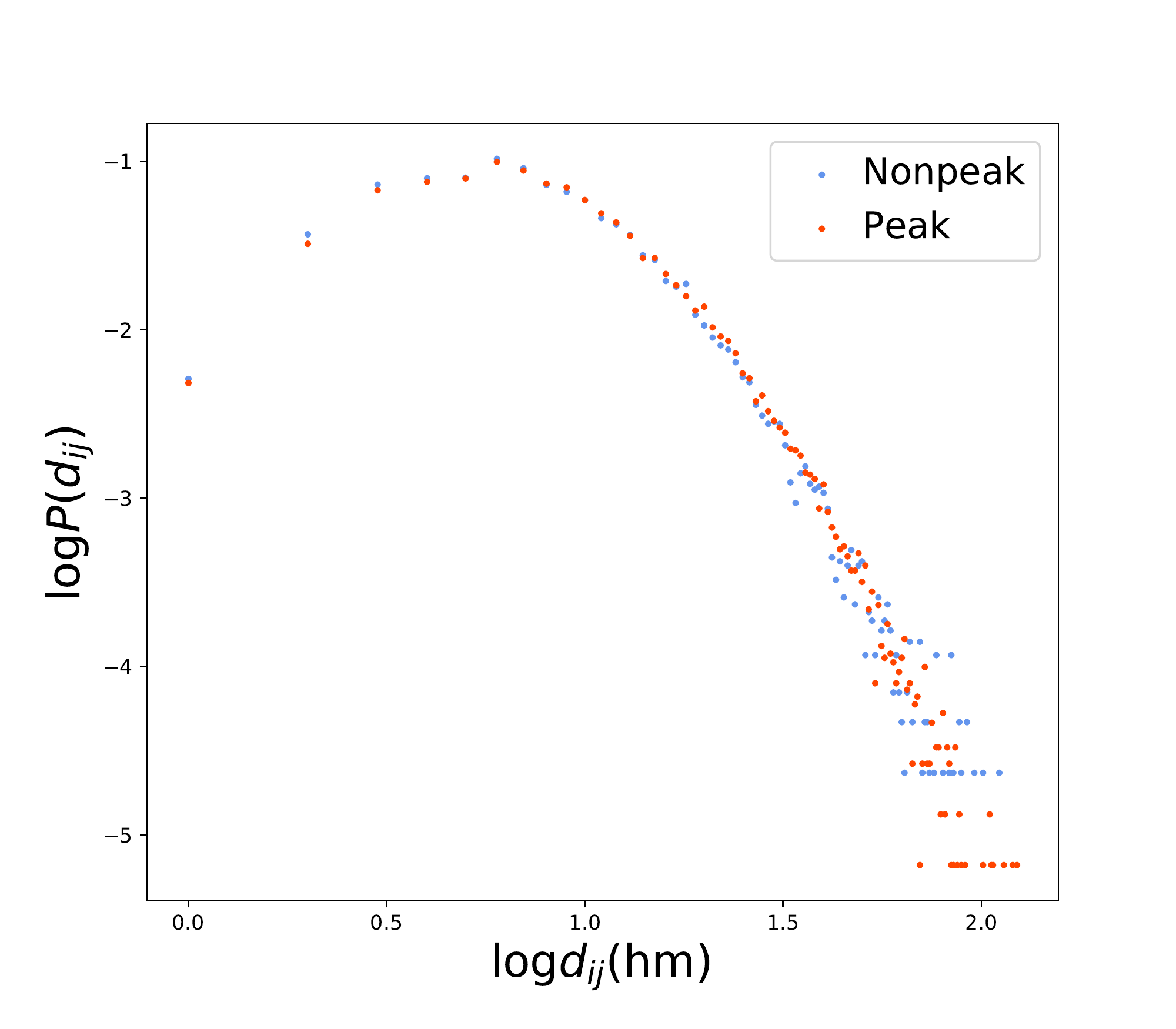}
    \end{minipage}}
\caption{Distribution of Euclidean cycling distance in three cities. }
\label{Figure 1}
\end{figure}

\section{Results}

\subsection{Data and preprocessing} 

In this study, we collect millions of cycling trips by dockless sharing bikes in three diversified cities in China: Beijing, Shanghai, and Xiamen. 
The Beijing dataset has 3.1 million trip records for roughly two weeks from May 10th to May 23rd, 2017. There are 1.022 million trip records in the Shanghai dataset from Aug. 1st to Aug. 31st, 2016. In the Xiamen dataset, there are 220,000 records from Dec. 21st to Dec. 25th, 2020. 
The Beijing and Shanghai datasets are obtained from Mobike platform, and the Xiamen dataset is aggregated from three largest platforms in the city -- Meituan Bike (formerly Mobike), Hellobike, and DiDi Bike.  
All datasets are formatted as follows: an order ID, an anonymized user ID, a bike ID, a start time and end time, a start position and an end position in (latitude, longitude). Different from Beijing and Shanghai datasets, which contain the data for the whole day, the Xiamen dataset is limited to 6 a.m. to 10 a.m. of each day and does not have the user ID for each record due to privacy and business concerns. 
To make a better traffic analysis, we rasterize the urban space into 500m$\times$500m grids (i.e., locations) for Beijing and Shanghai and 300m$\times$300m grids for Xiamen, as Beijing and Shanghai are relatively large and Xiamen is a small island with a diameter of less than ten miles. Then we associate each start and end position to certain locations. 

For the raw data, we do some simple filtering: we discard the records with origin or destination not located within the boundary of the city and also discard the ones with a riding duration longer than one day (which might be some bikes left unlocked after their initial order) or less than one minute (which might be due to unsatisfied tryouts). 
To avoid possible biases, we do not pose any further filtering criteria. 

The Euclidean distance of trips by dockless sharing bikes in cities can be well approximated by a log-normal distribution (see Fig. \ref{Figure 1}), which peaks at around 1 km but has a non-negligible fraction of trips with much longer displacements regardless of during rush hours or off-peak hours. 
And it is worth noting that there are more trips with a longer distance during rush hours than during off-peak hours in both Beijing and Shanghai. Such a pattern is different from the taxi, which tends to have a shorter trip distance and duration during rush hours than during off-peak hours \cite{feng2022scaling}. This indicates that for commuting, people may have to ride bicycles for longer distances to save time (see Fig. \ref{Figure 2}) or to avoid crowding in public transportation, and this happens mainly for trips roughly longer than two kilometers and shorter than ten kilometers. In addition, some recent evidence also indicates that cycling is of the highest quality of commuting experience when compared to driving, taking buses, and walking \cite{liu2021exploring}. While during off-peak hours, when trips are not that urgent (e.g., sightseeing or exercise), people have a weaker tendency to cycle for relatively long trips (see Fig. \ref{Figure 1}). 
Here, peak hours are from 7 a.m. to 9 a.m. and 5 p.m. to 8 p.m., and other time slots are off-peak hours. As the Xiamen dataset is from 6 a.m. to 10 a.m., thus results shown in Fig. \ref{Figure 3}(c) are limited to this time period, which also explains why distribution of displacement distance is similar there. 
From the patterns shown in Fig. \ref{Figure 1} and Fig. \ref{Figure 2}, we speculate that people use dockless sharing bikes mainly for short trips to get around the neighborhood quickly \cite{fishman2013bike,noland2006smart,shaheen2013public} or to connect commuting \cite{chen2019analyzing}, but also ride them for long trips relatively frequently than a normal or Poisson distribution would assume. 
Some evidence indicates that users at the fringe of cities usually have a long average weighted travel distance by dockless sharing bikes, which indicates that sharing bikes serve as more than just short-distance commuting connections there \cite{li2021gravity} or a less developed public transportation there. 

\begin{figure}    \centering
\subfigure[Beijing]
{    \begin{minipage}[b]{.3\linewidth}        \centering
        \includegraphics[width=\linewidth]{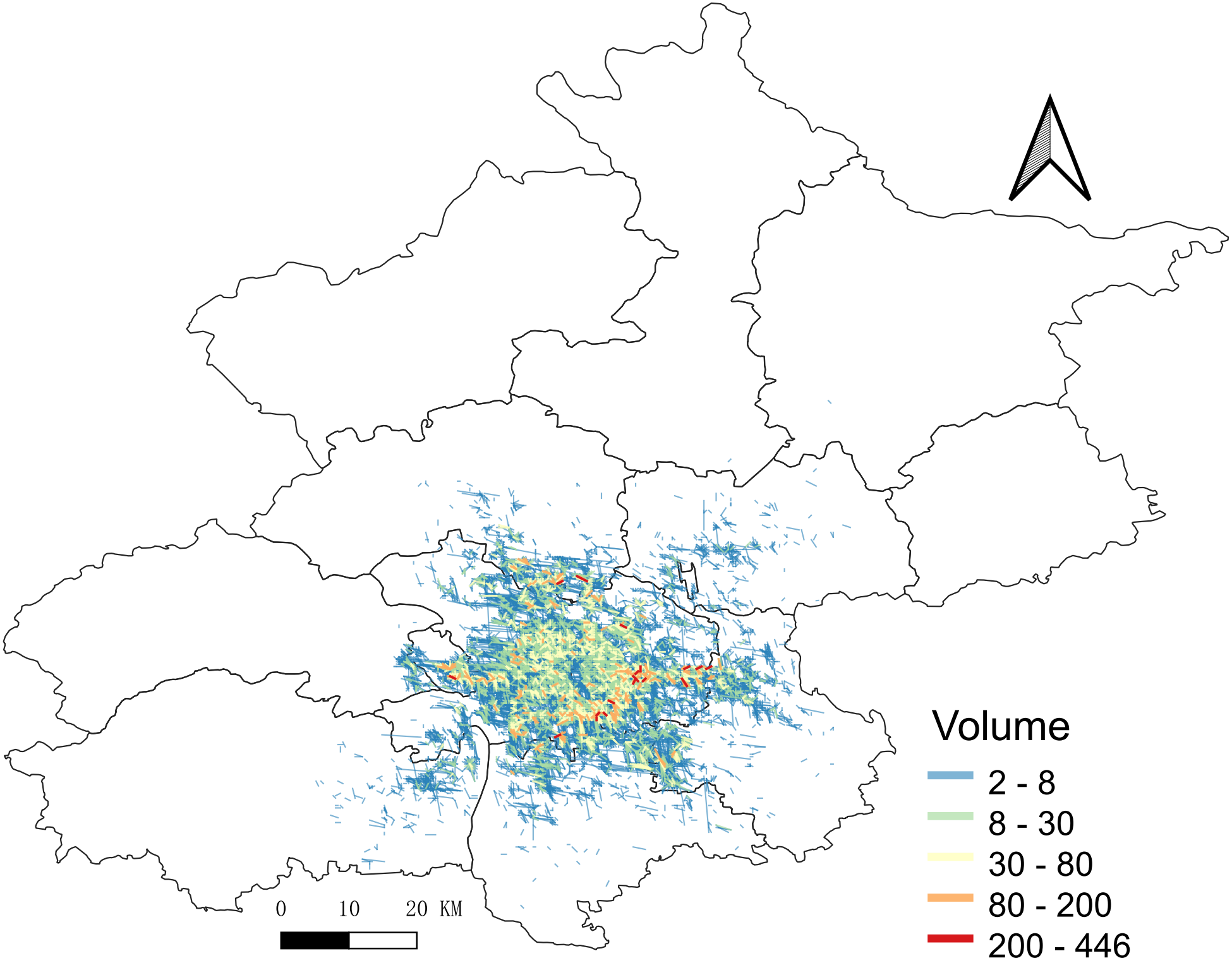}
    \end{minipage}}
\subfigure[Shanghai]
{ 	\begin{minipage}[b]{.3\linewidth}        \centering
        \includegraphics[width=0.85\linewidth]{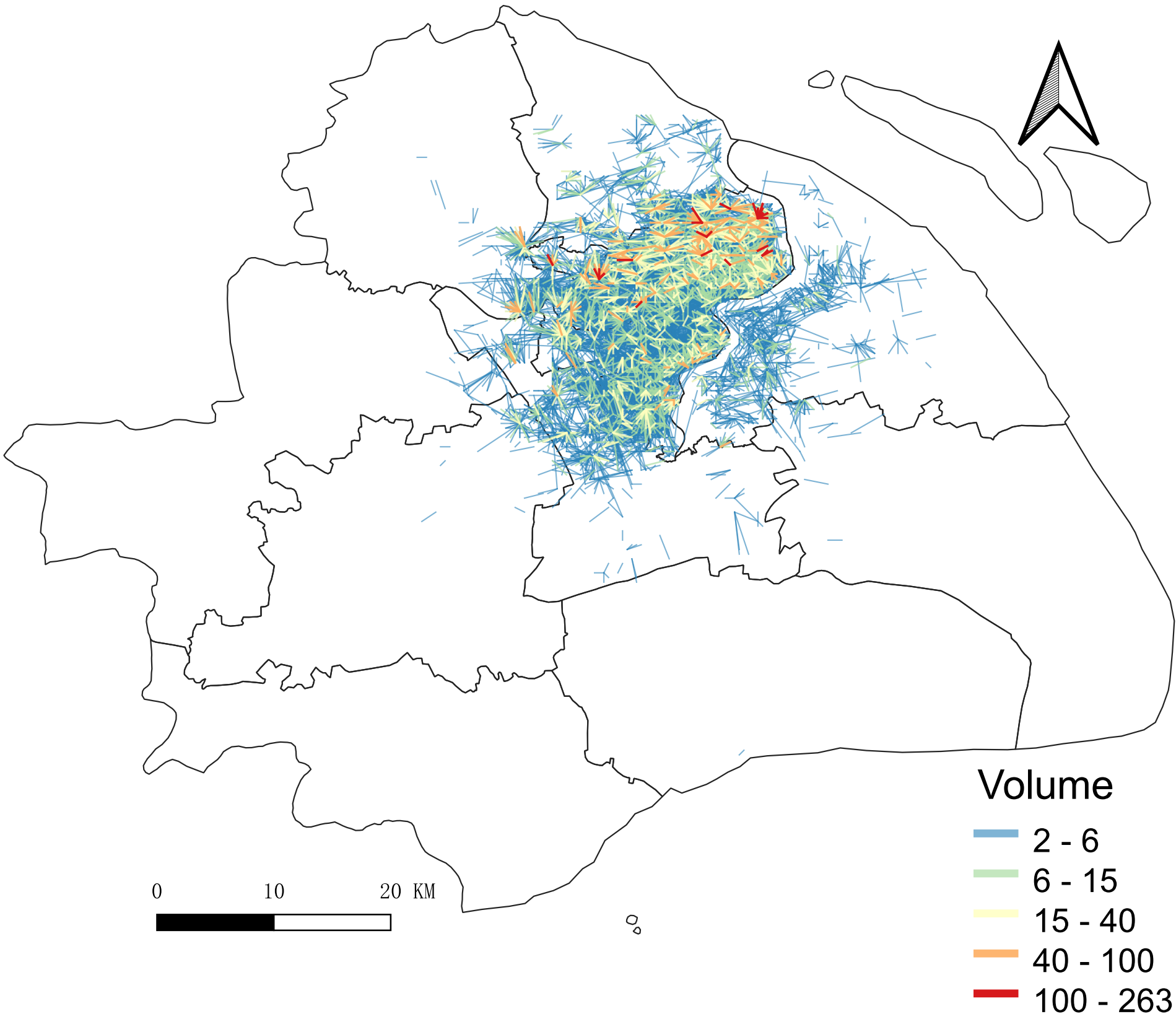}
    \end{minipage}}
\subfigure[Xiamen]
{ 	\begin{minipage}[b]{.3\linewidth}        \centering
        \includegraphics[width=0.85\linewidth]{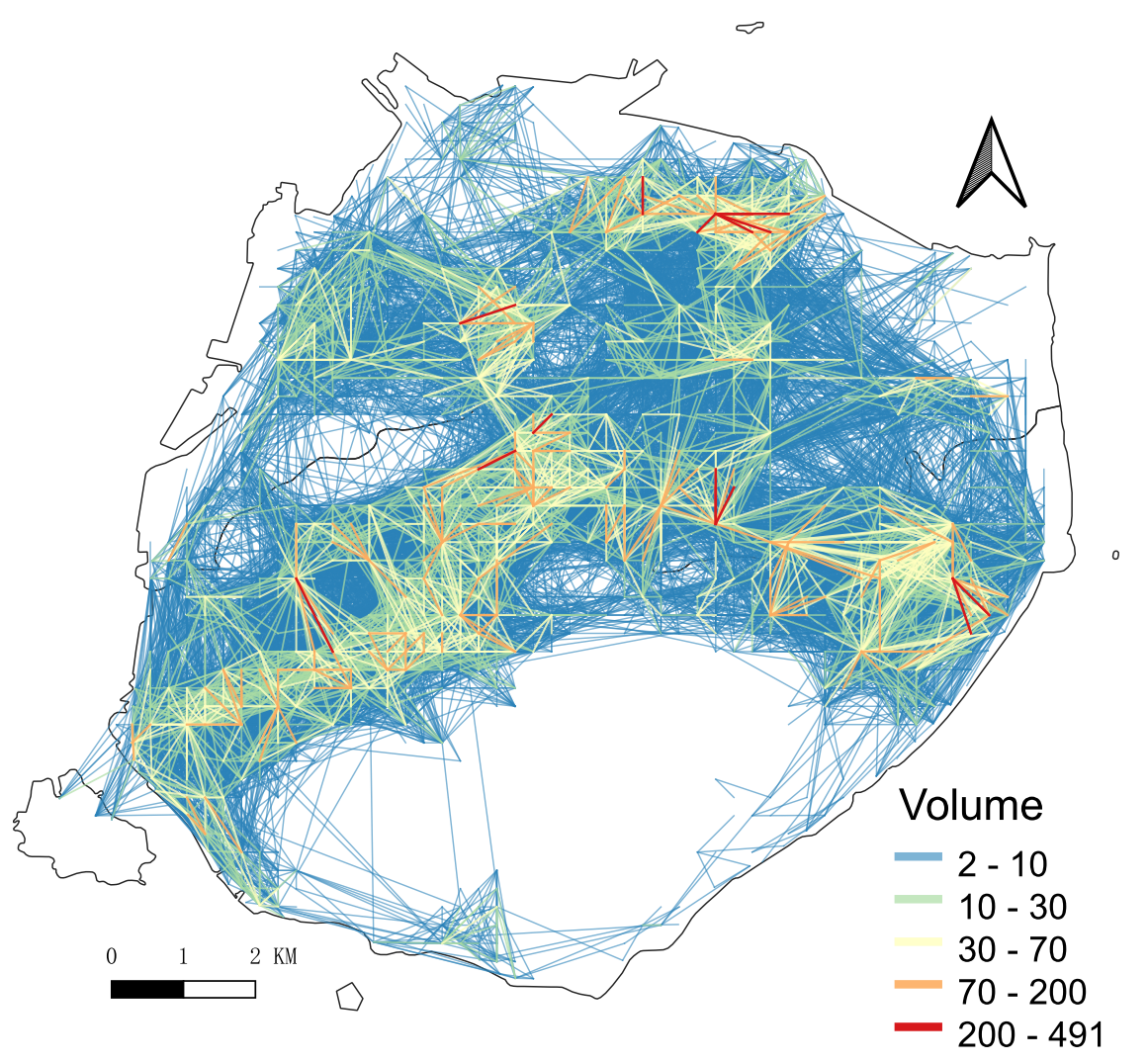}
    \end{minipage}}
\subfigure[Beijing]
{    \begin{minipage}[b]{.3\linewidth}        \centering
        \includegraphics[width=\linewidth]{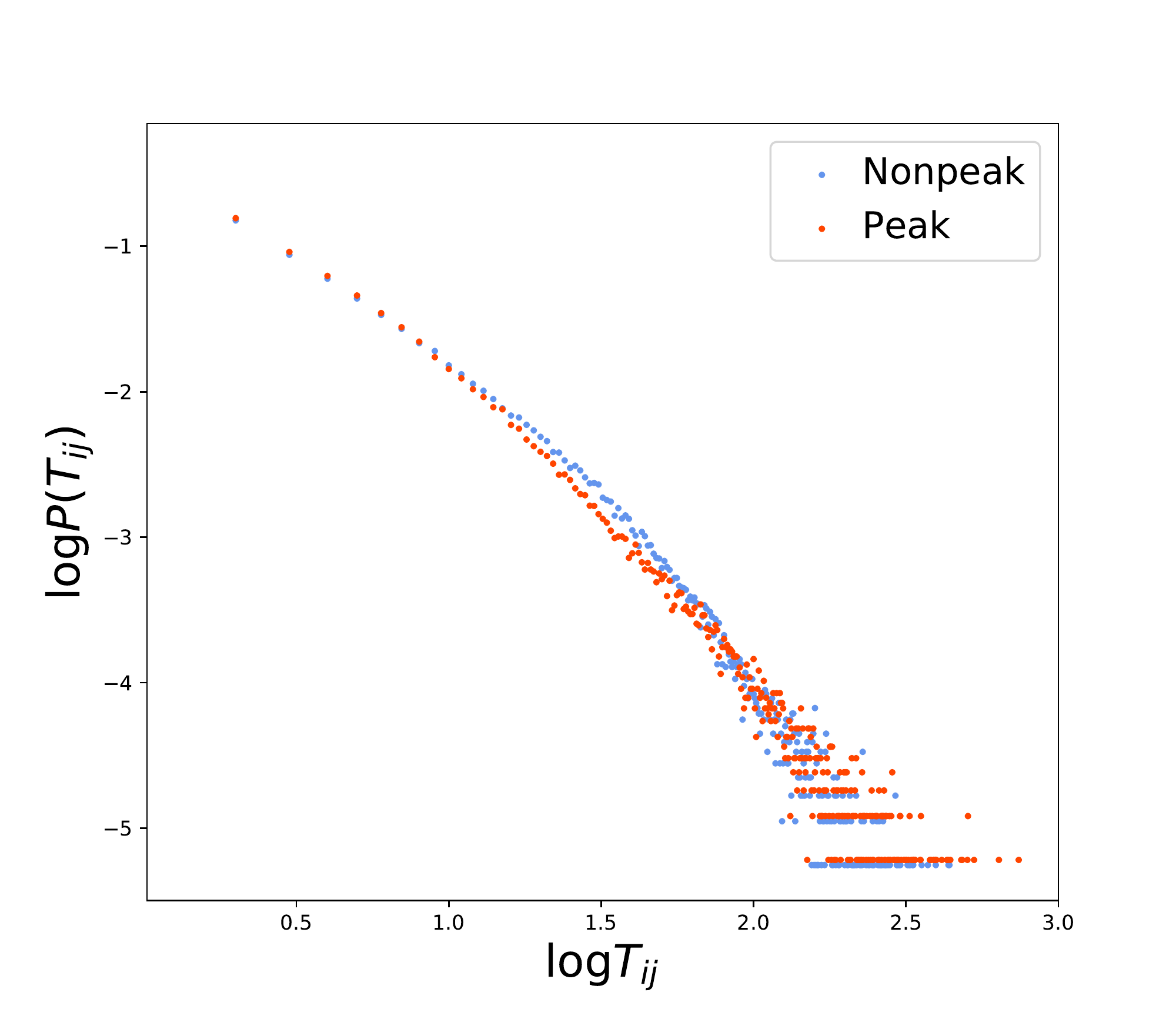}
    \end{minipage}}
\subfigure[Shanghai]
{ 	\begin{minipage}[b]{.3\linewidth}        \centering
        \includegraphics[width=\linewidth]{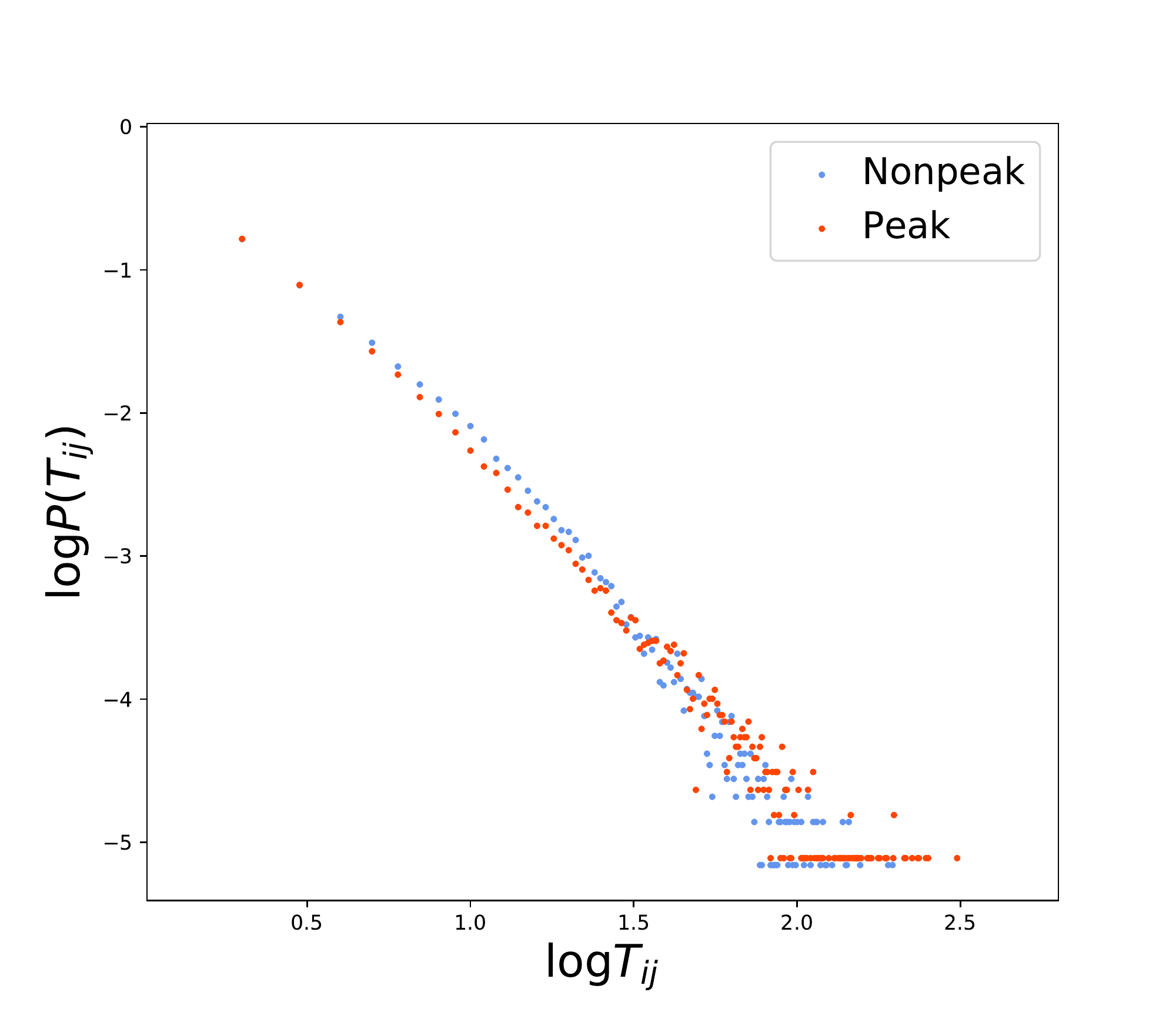}
    \end{minipage}}
\subfigure[Xiamen]
{ 	\begin{minipage}[b]{.3\linewidth}        \centering
        \includegraphics[width=\linewidth]{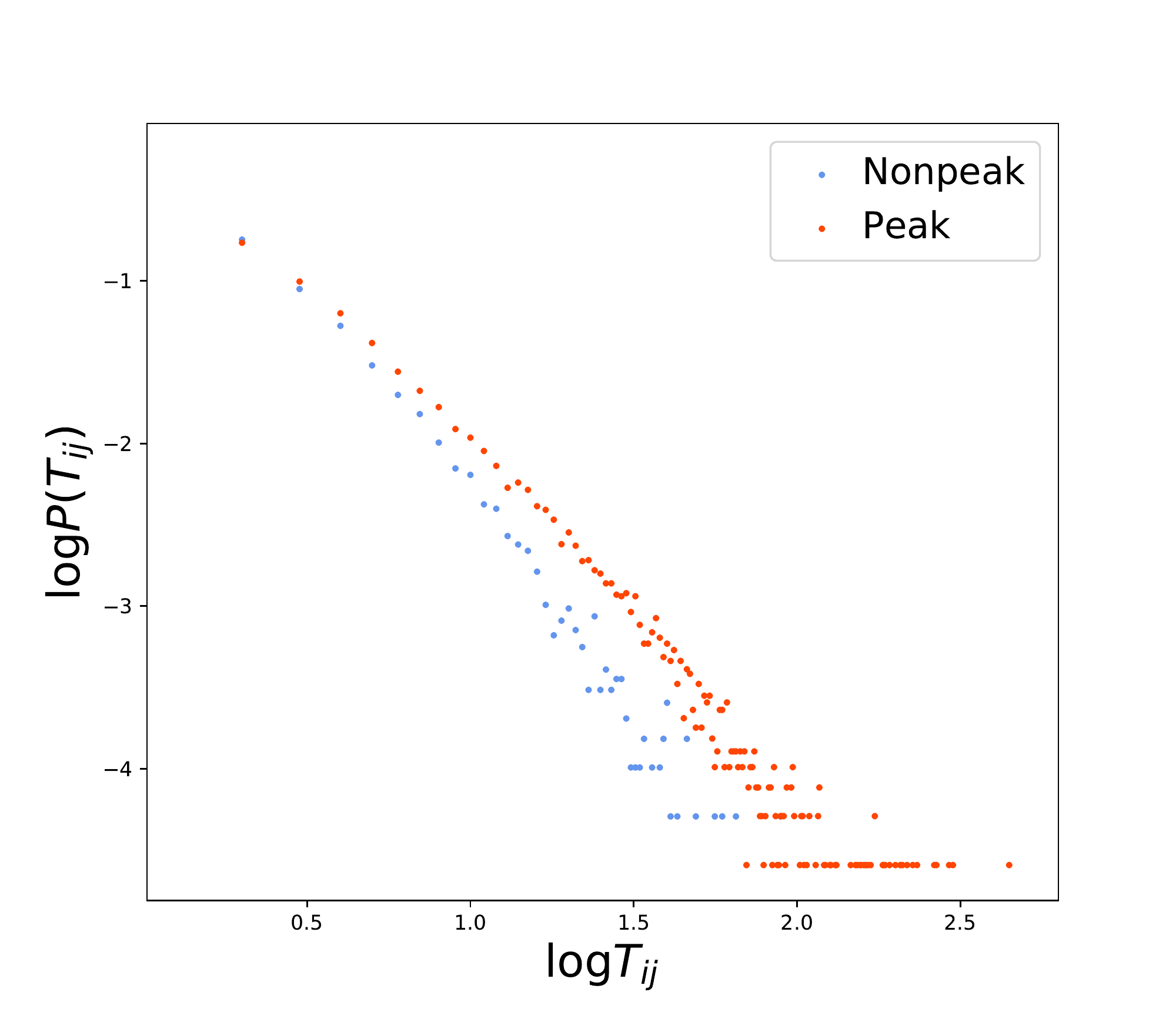}
    \end{minipage}}
    \caption{Distribution of biking traffic in cities. For better clarity, biking traffic between location pairs that equals one is not shown in (a)-(c).}
    \label{fig.bikingtraffic}
\end{figure}

When aggregating each trip to locations, we obtain the origin-destination matrix of biking traffic in cities (see Fig. \ref{fig.bikingtraffic}). Cycling traffic in the peripheral regions of the city is generally low, but location pairs generating larger biking flow are also relatively dispersed, which are near dense residential communities and workplaces. 
Apart from being influenced by population \cite{li2021gravity}, bicycle traffic is also constrained by geographical factors. For example, there are quite few trips crossing Huangpu River in Shanghai (see Fig. \ref{fig.bikingtraffic}(b)) or near Xiamen Botanical Garden and the area of Dongping mountain (see Fig. \ref{fig.bikingtraffic}(c)).
Despite varying geographical characteristics across cities, the distributions of biking traffic all follow a power-law during both rush hours and off-peak hours (see Fig. \ref{fig.bikingtraffic}(d)-(f)). 
This finding indicates a strong heterogeneity of biking flows that most routes have low bicycle traffic but there is also a notable fraction of routes has high bicycle traffic which takes up a large fraction of biking traffic in the whole city (see Fig. \ref{fig.bikingtraffic}(a)-(c)). 

\subsection{Road network topology and geographical characteristics of the cycling section}

\begin{figure}[htbp] \centering
\subfigure[Beijing]
{    \begin{minipage}[b]{.3\linewidth}        \centering
        \includegraphics[scale=0.25]{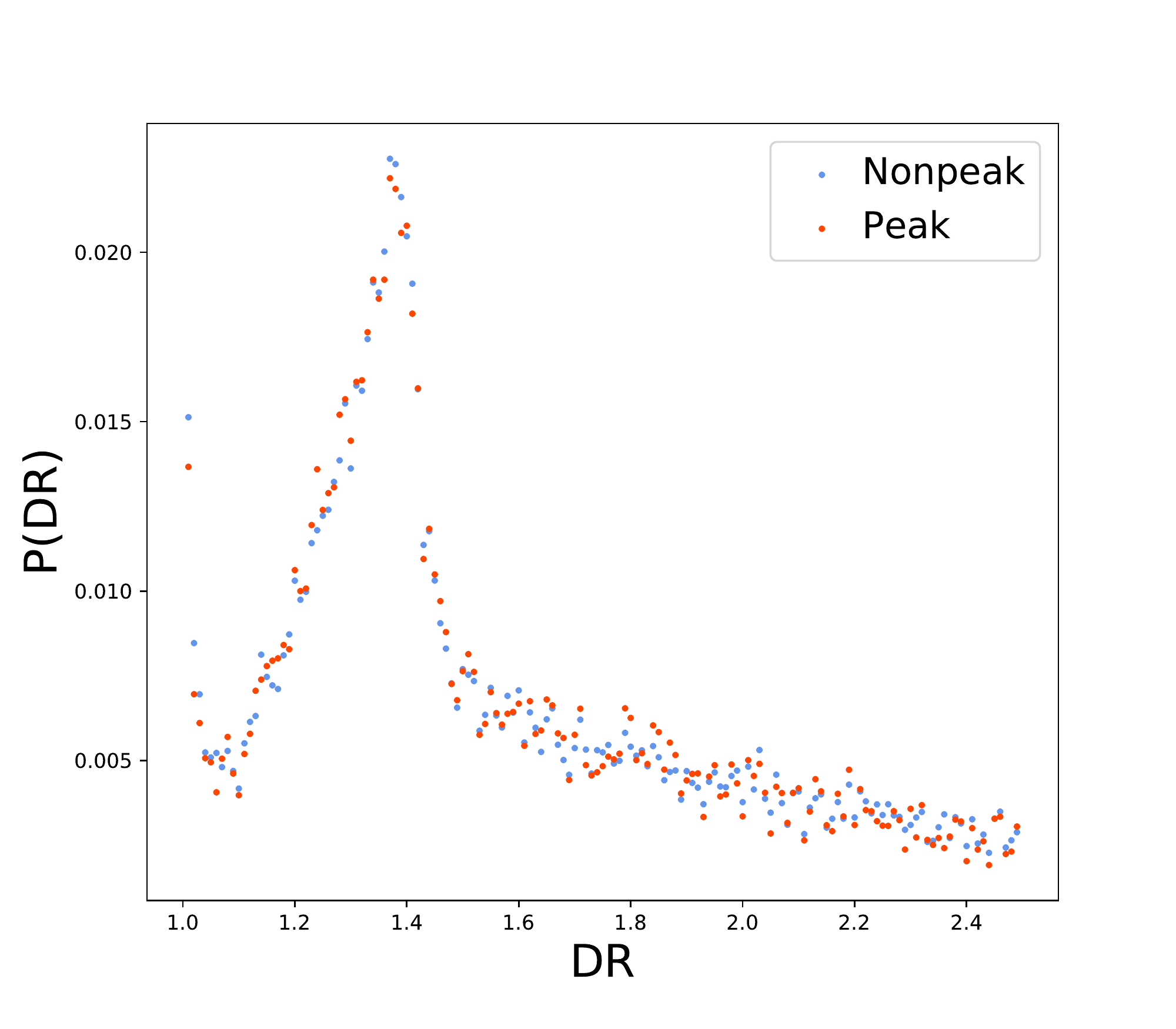}
    \end{minipage}}
\subfigure[Shanghai]
{ 	\begin{minipage}[b]{.3\linewidth}        \centering
        \includegraphics[scale=0.25]{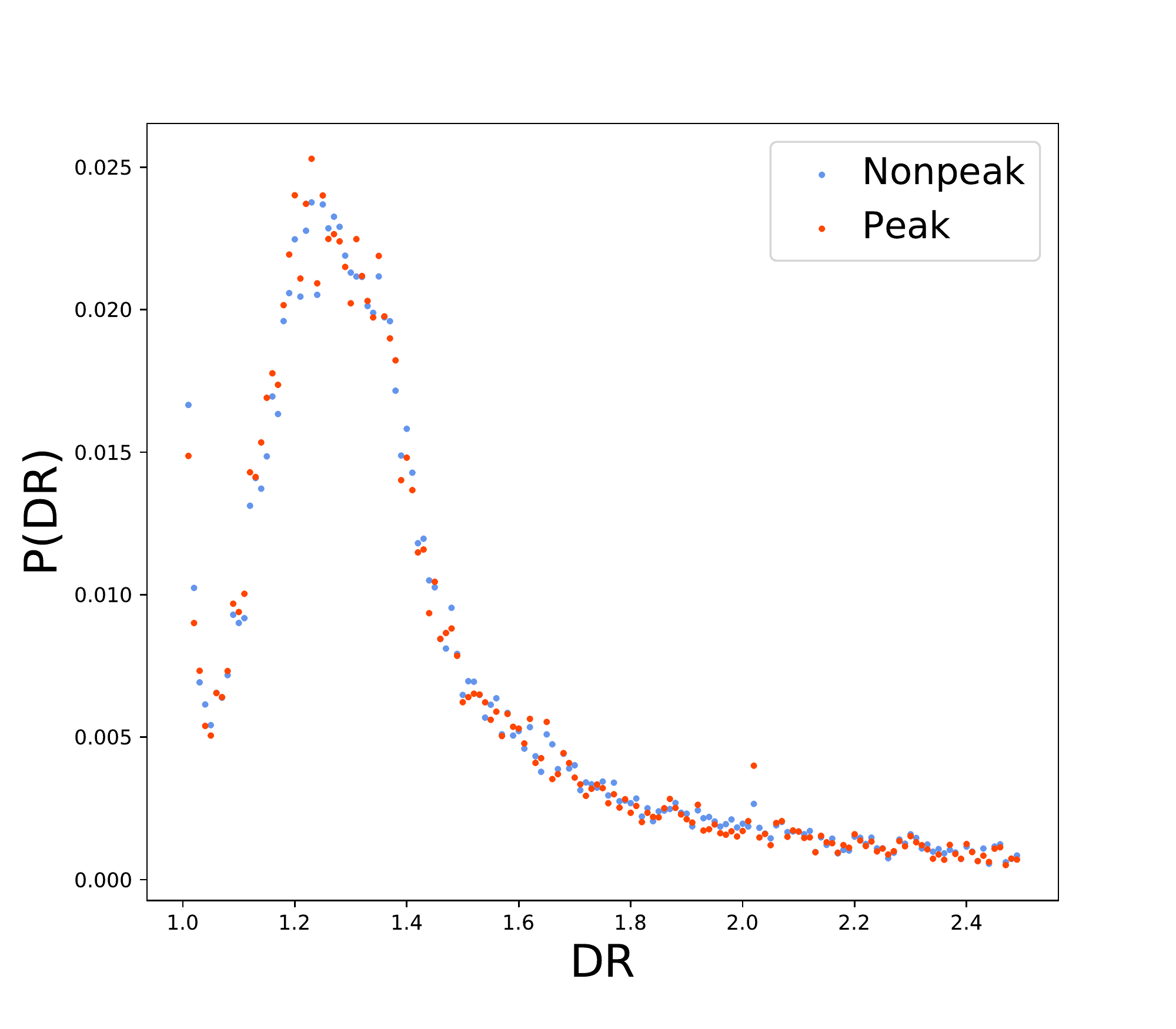}
    \end{minipage}}
\subfigure[Xiamen]
{ 	\begin{minipage}[b]{.3\linewidth}        \centering
        \includegraphics[scale=0.25]{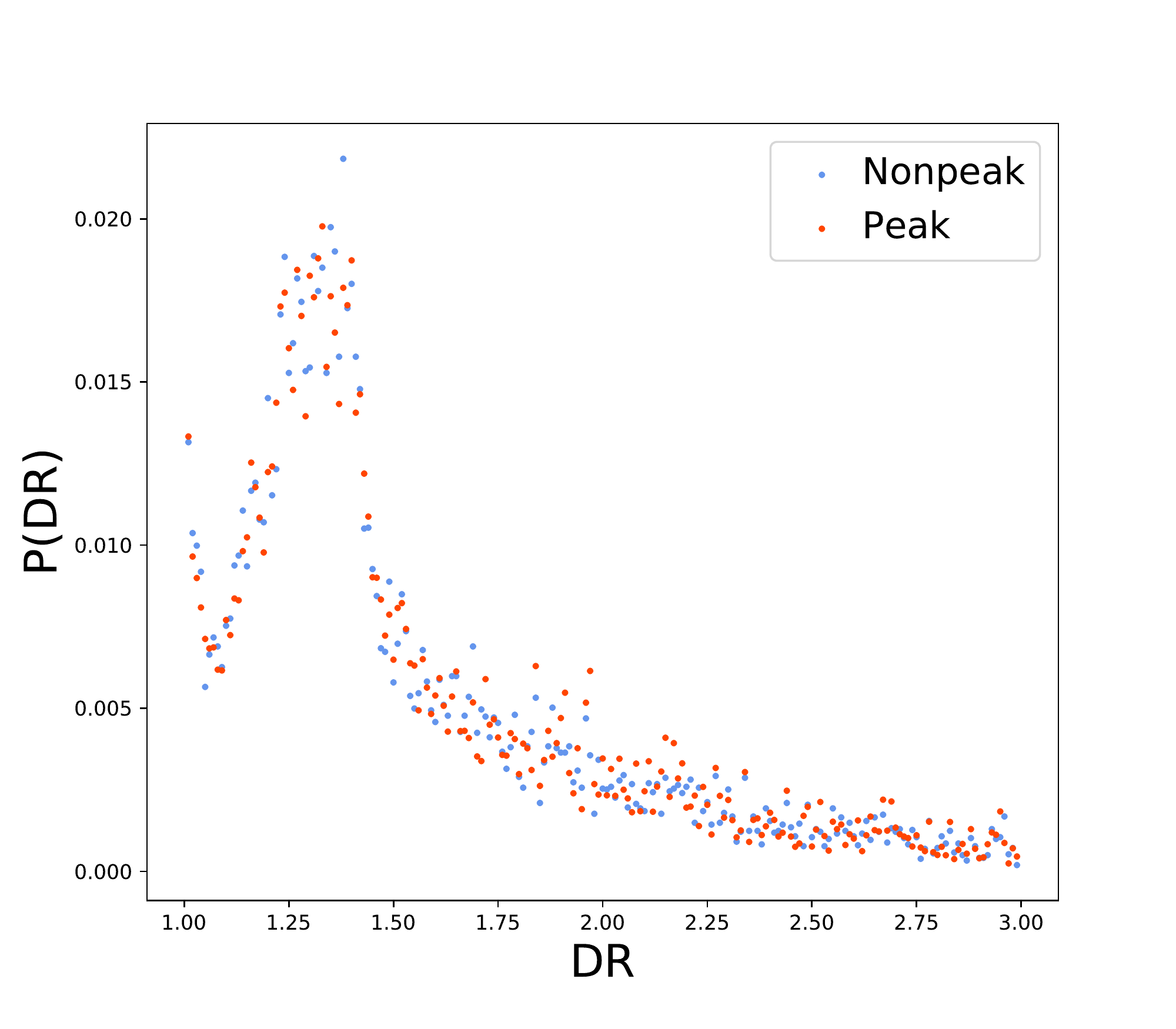}
    \end{minipage}}
\caption{Distribution of detour ratio for biking trips in three cities} 
\label{Figure 3}
\end{figure}

In traffic engineering and geographical analysis of human activities, the detour ratio (DR) \cite{black2003transportation,gastner2006shape,barthelemy2011spatial,dong2016population,yang2018universal} has been widely applied for assessing transportation efficiency, and recent advances also indicate a strong relation with navigation cognitive ability of residents \cite{coutrot2022entropy}, which has great impacts on both cycling activities and vehicle traffic. The DR equals the actual routing distance $r_{ij}$ on the road network over the Euclidean distance $d_{ij}$ between two locations $i$ and $j$,  
which is formulated as
\begin{equation}
    DR_{ij} = \frac{r_{ij}}{d_{ij}},
\end{equation}
where  
$r_{ij}$ is queried from Amap API (\url{https://lbs.amap.com}) for the fastest (usually shortest) routing path for each trip by car, $d_{ij}$ is from the great-circle distance between the start and end locations, and the value of $DR_{ij}$ is larger or equal to one. Here, we only consider routes with biking traffic.  
The distribution of DR for location pairs with cycling activities during both peak hours and off-peak hours are almost identical (see Fig. \ref{Figure 3}), which indicates that spatial patterns of cycling trips are quite similar during different times. 

In general, DR peaks around 1.3 to 1.4 and exhibits similar distribution across cities, which further reflects 
universal regularities behind the interplay between topology and geographic structure of road networks over diversified cities \cite{yang2018universal}.  
In addition, we can observe that distributions of DR in Beijing and Xiamen are more right-skewed than that in Shanghai and Xiamen has more routes with a larger DR, which might be due to a larger block size \cite{louf2014typology} and segmenting effects posed by wider roads (e.g., 2nd to 5th Ring Roads are highways located within the urban area of Beijing) or a more complex urban terrain (e.g., several mountains and lakes are located in the urban area of Xiamen).  
In Chinese cities, gated communities \cite{li2021heightened} and working units and universities generally make trips more detoured, as outside traffic is usually unable to pass through and has to detour when encountering such enclosed regions.  
In Beijing, the block size, as well as road width (see Fig. \ref{fig:Road width}), is usually larger than in other cities. 
The existence of Ring Roads and other wide roads would also highly probably cause longer detours when there is no pass-through nearby. 
It is worth noting that the improvement priority of detoured routes should also depend on travel demands on them \cite{dong2016population}. If a very detoured route is used by many people for commuting, then it should be improved with a higher priority; while, if quite a few people are using it, then it will have a milder impact. 

\begin{figure}[!htbp] \centering
\subfigure[Beijing]
{    \begin{minipage}[b]{.3\linewidth}        \centering        \includegraphics[scale=0.25]{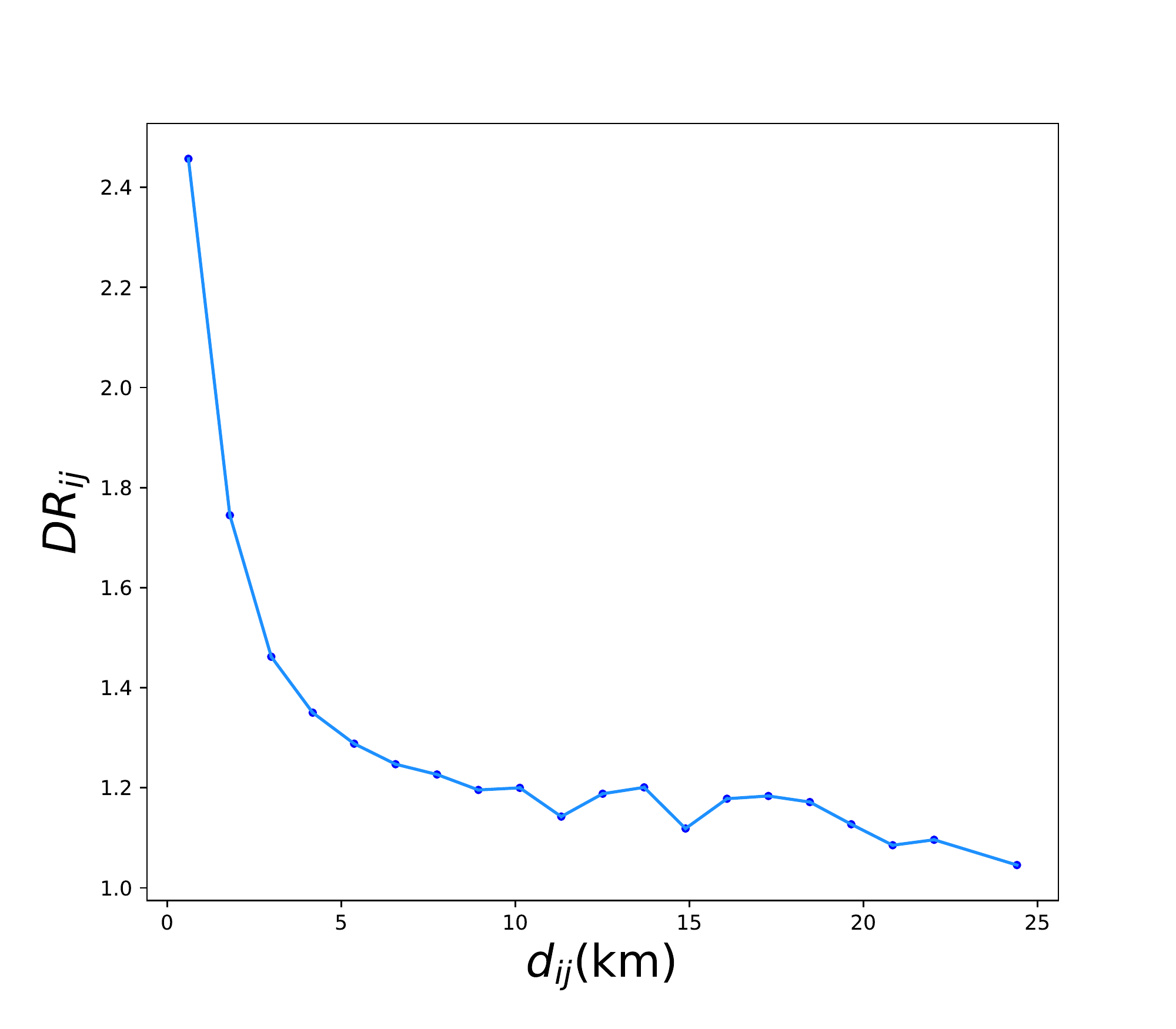}
    \end{minipage}}
\subfigure[Shanghai]
{ 	\begin{minipage}[b]{.3\linewidth}        \centering
        \includegraphics[scale=0.25]{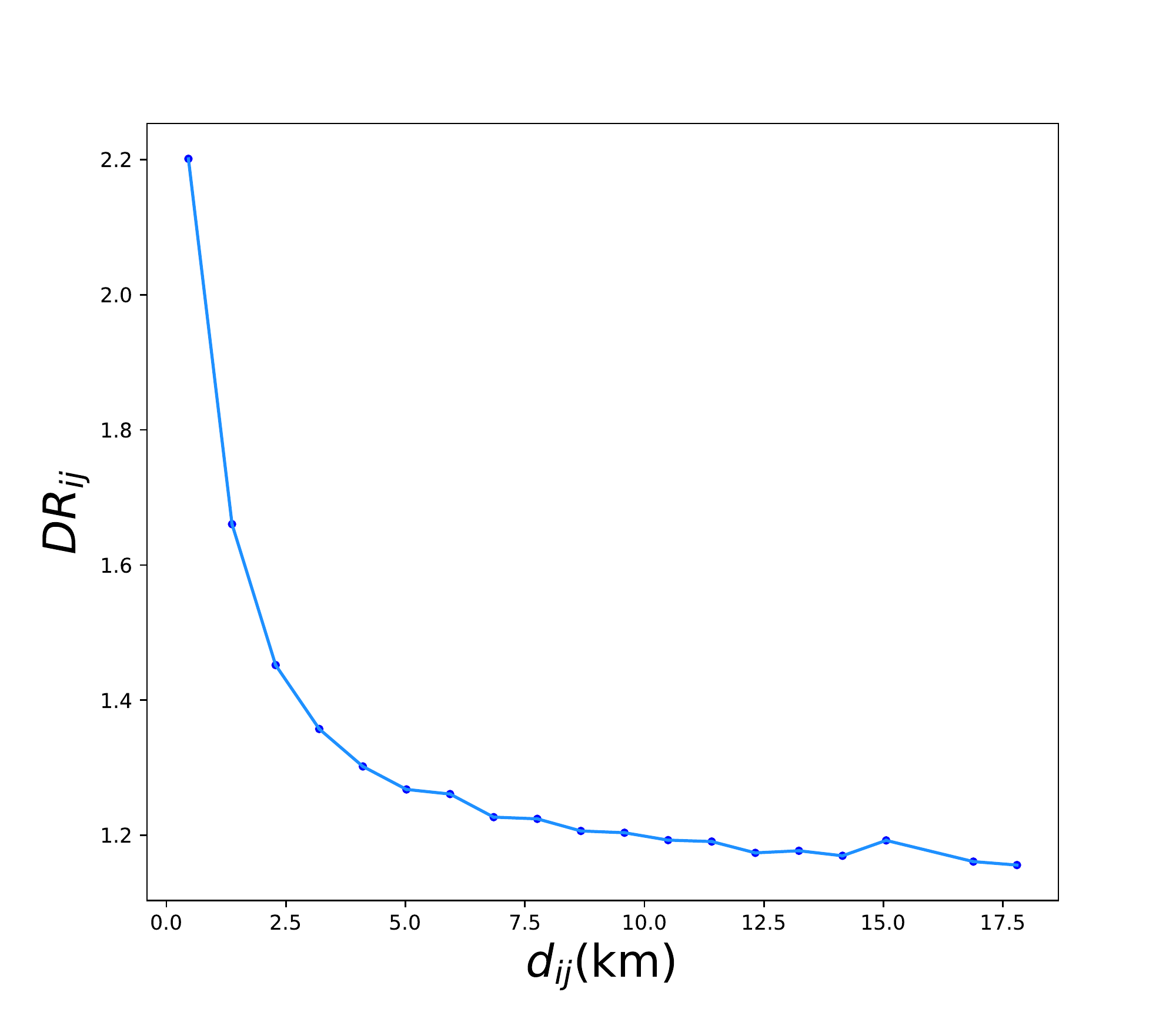}
    \end{minipage}}
\subfigure[Xiamen]
{ 	\begin{minipage}[b]{.3\linewidth}        \centering
        \includegraphics[scale=0.25]{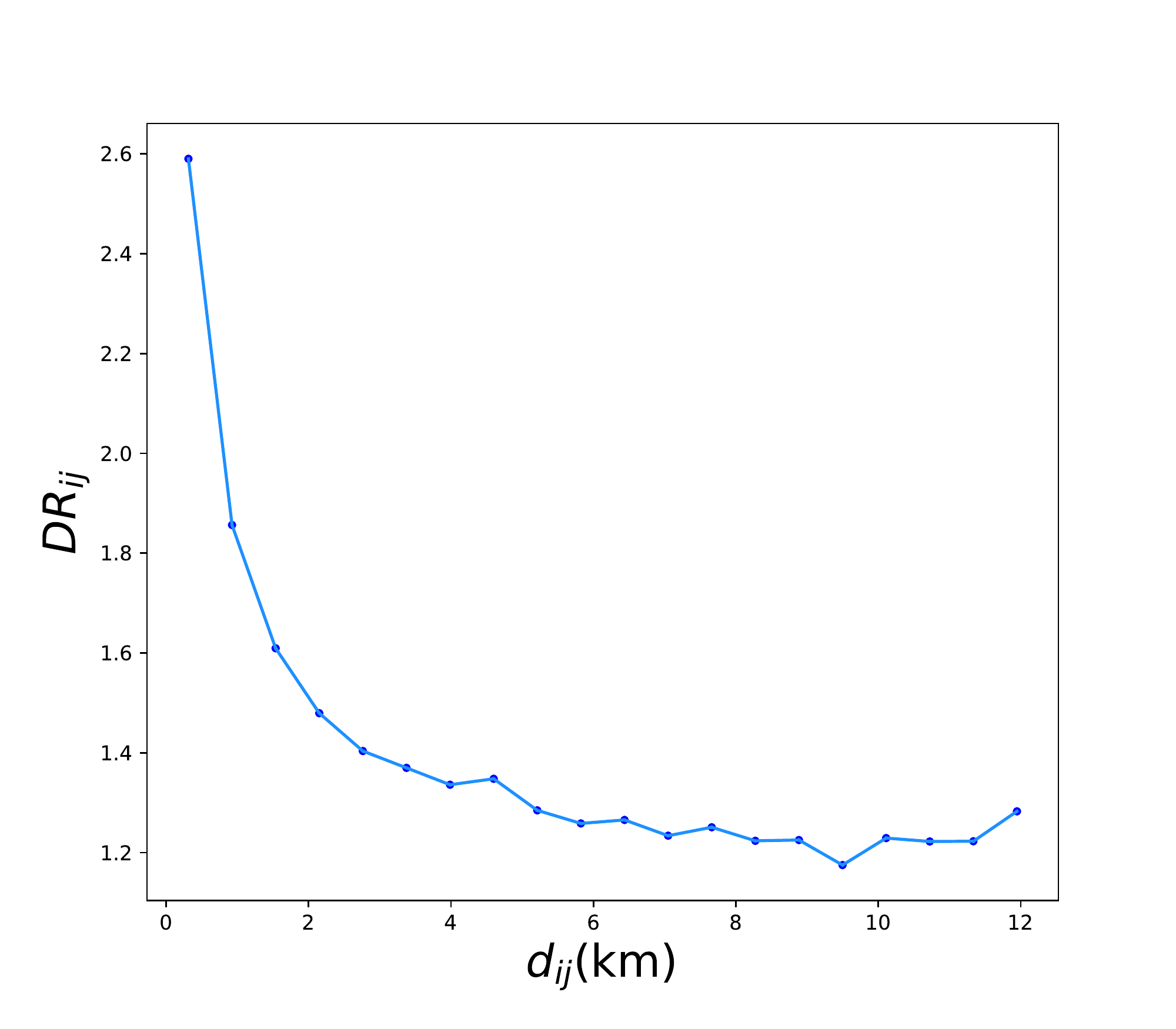}
    \end{minipage}}
\caption{Detour ratio against Euclidean distance for biking trips in three cities}
\label{Figure 4}
\end{figure}

We find that the DR shows a nonlinear negative correlation with the Euclidean distance in all three cities, 
and most roads with large DR generally have Euclidean distance less than a few kilometers. 
The existence of one-way roads \cite{yang2018universal}, where people can only drive or cycle in one direction, would also result in a higher DR. In practice, bidirectional biking traffic may exist even on one-way roads, and this would usually lead to disturbances to pedestrians or normal vehicle traffic due to a higher chance of dangerous encountering between vehicles and bikes.

 \begin{figure}[htbp]\centering
 \subfigure[Beijing]{
 \begin{minipage}[t]{.3\linewidth}\centering
 \includegraphics[scale=0.25]{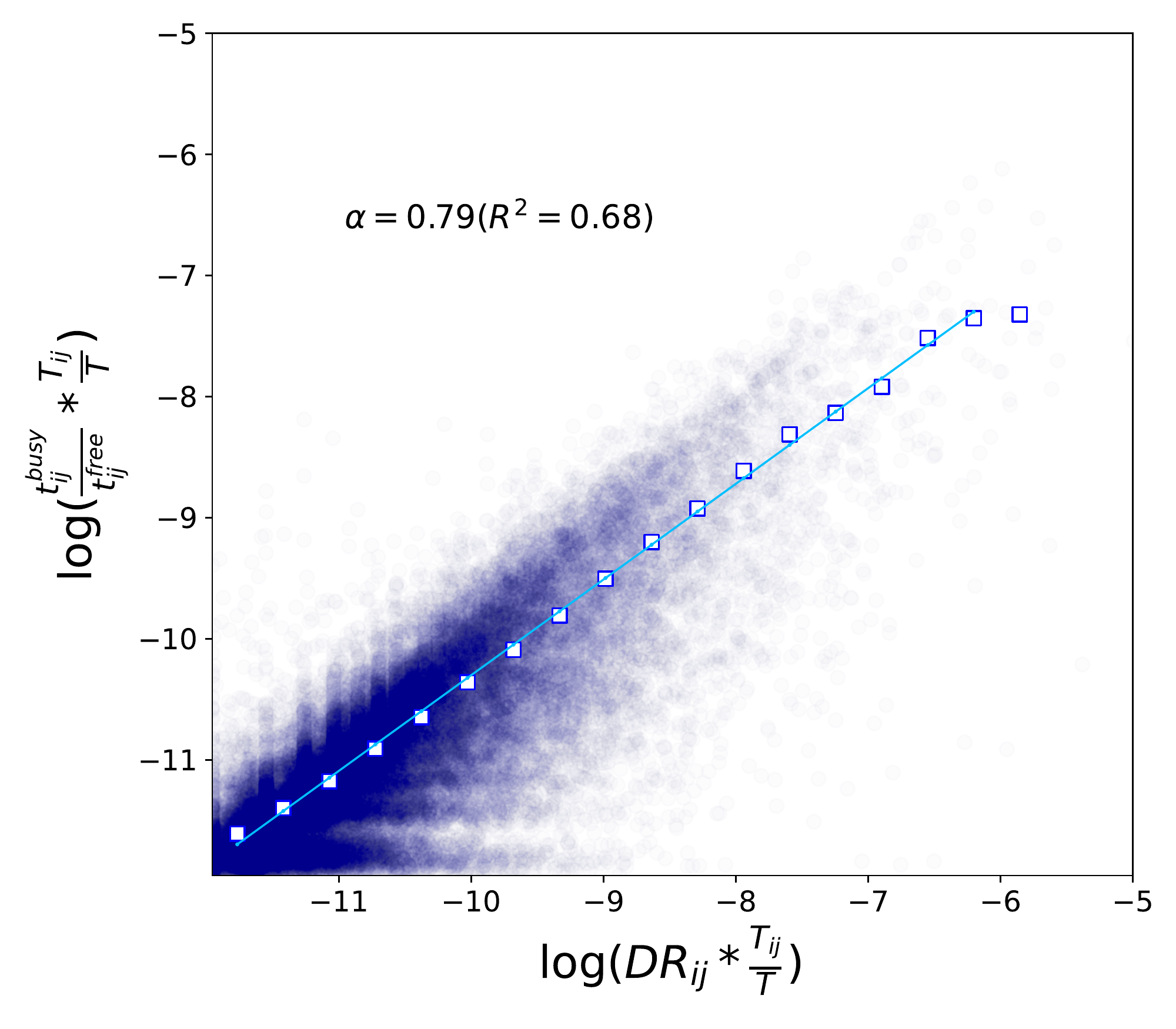}
 \end{minipage}}%
 \subfigure[Shanghai]{
 \begin{minipage}[t]{.3\linewidth}\centering
 \includegraphics[scale=0.25]{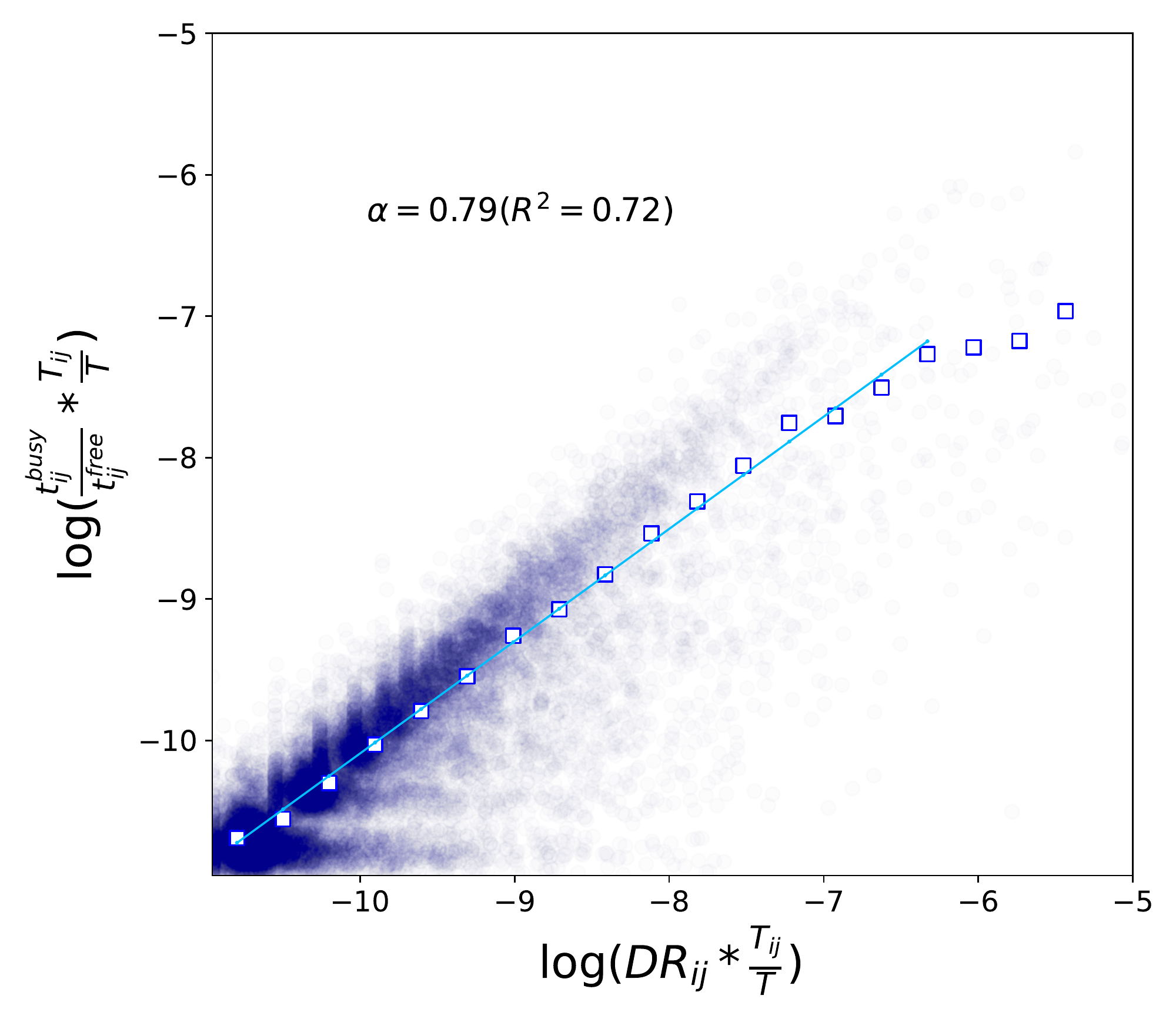}
 \end{minipage}}%
 \subfigure[Xiamen]{
 \begin{minipage}[t]{.3\linewidth}\centering
 \includegraphics[scale=0.25]{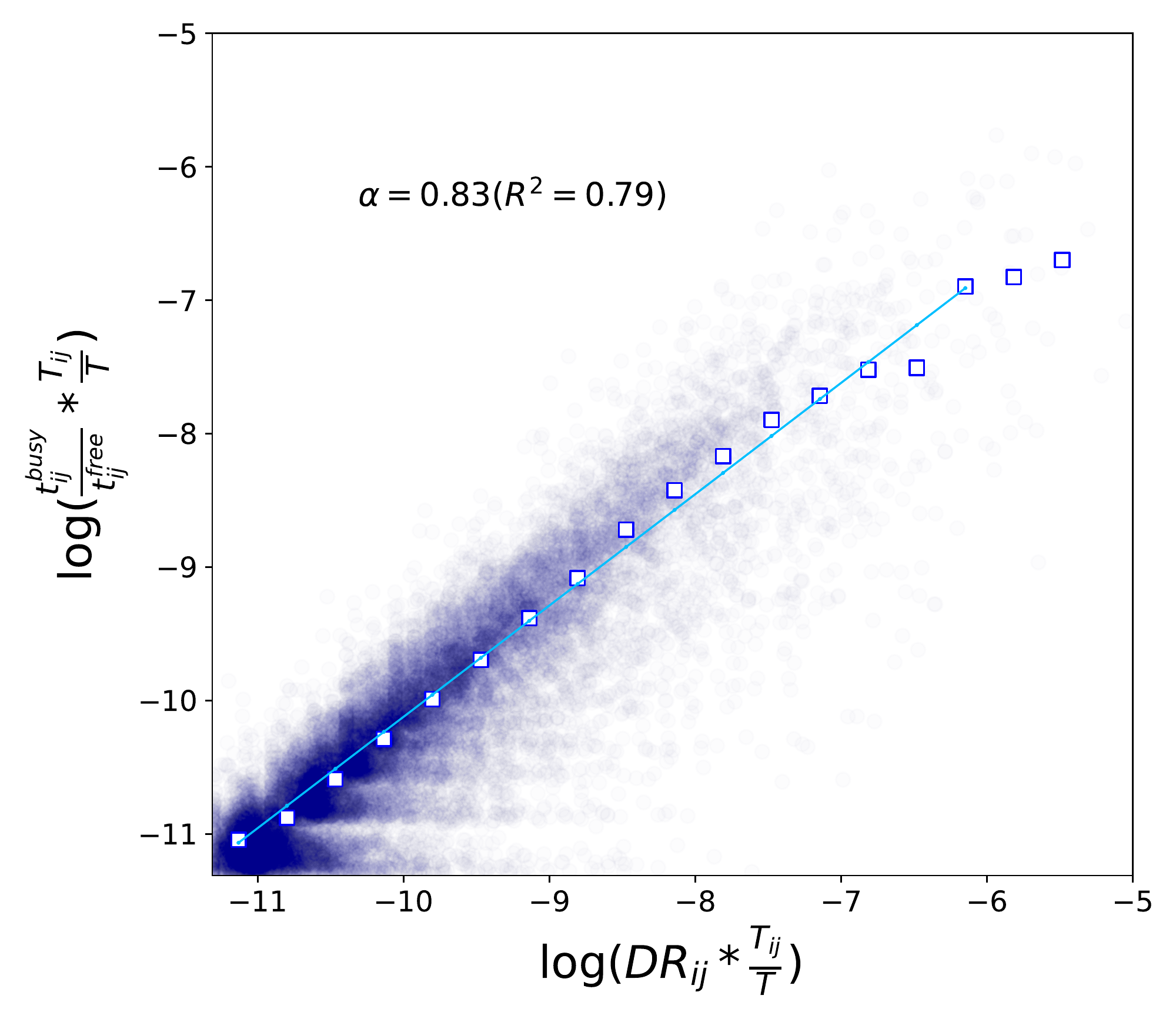}
 \end{minipage}}%
 \quad
 \subfigure[Beijing]{
 \begin{minipage}[t]{.3\linewidth}\centering
 \includegraphics[scale=0.25]{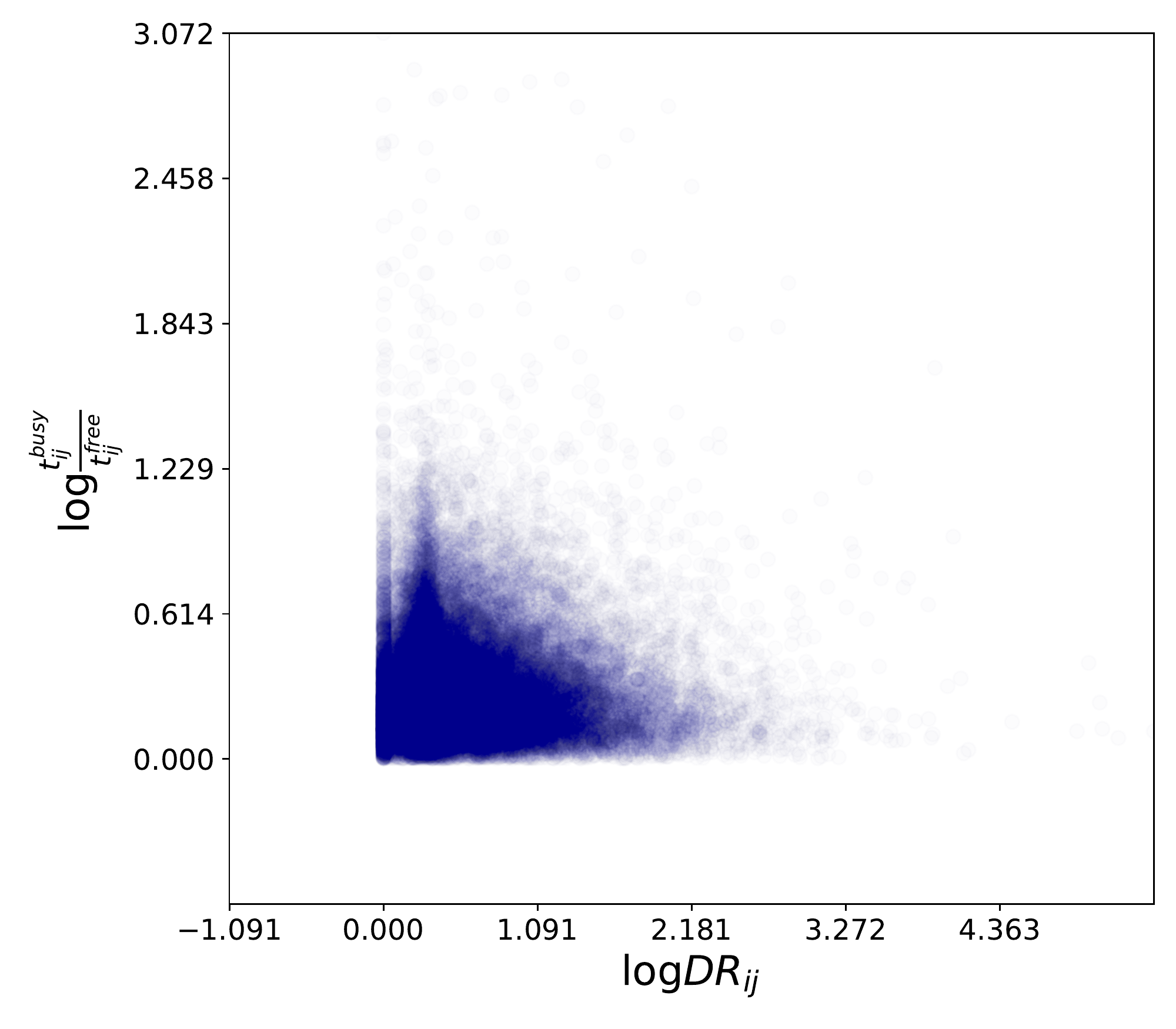}
 \end{minipage}}%
 \subfigure[Shanghai]{
 \begin{minipage}[t]{.3\linewidth}\centering
 \includegraphics[scale=0.25]{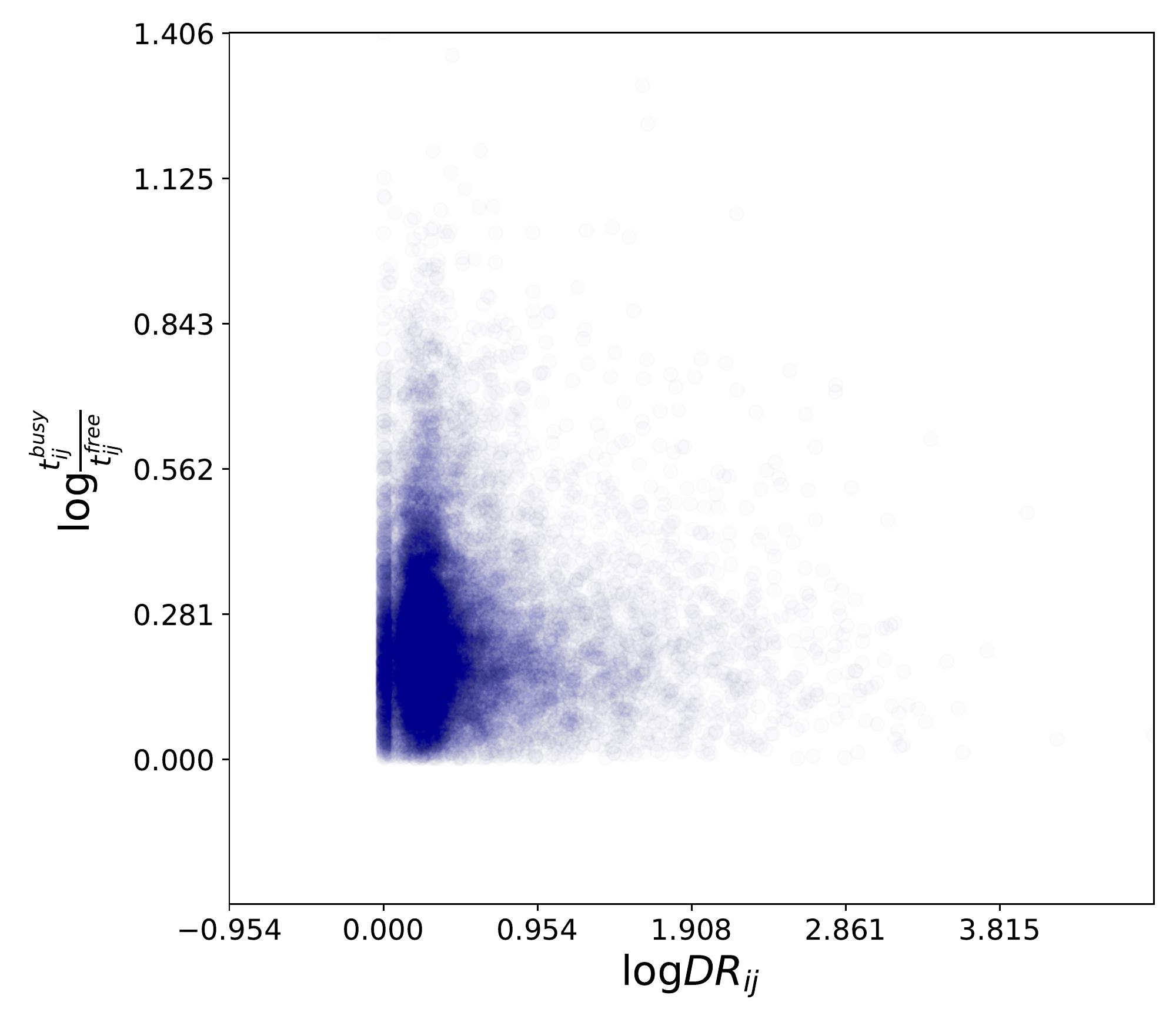}
 \end{minipage}}%
 \subfigure[Xiamen]{
 \begin{minipage}[t]{.3\linewidth}\centering
 \includegraphics[scale=0.25]{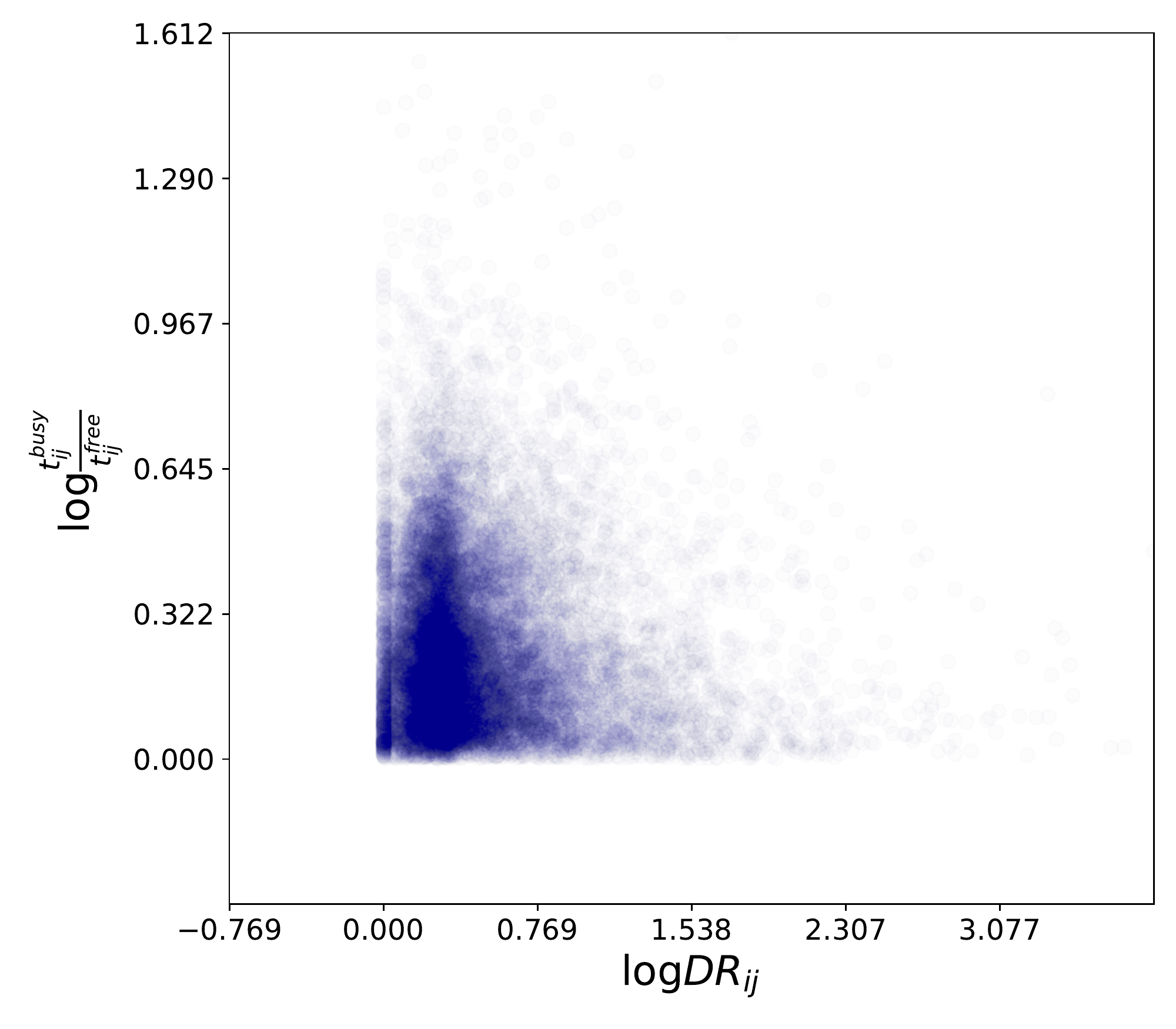}
 \end{minipage}}
 \caption{Congestion against biking-traffic weighted and unweighted DR}
 \label{Figure 5}
 \end{figure}

\subsection{Hidden interplay between travel modes and road networks}
To further analyze the relationship between biking traffic and road network structure and urban congestion, we mainly focus on commuting behaviors. As dockless sharing bikes are more like a ``first-mile facilitator'' in the morning rather than a ``last-mile solution'' in the evening \cite{xu2019unravelBIKE}, we extract the data during morning rush hours (6:00 a.m. to 10:00 a.m.) in working days in all three cities. From the Amap API, we further query the vehicle travel time (i.e., driving time) for each location pair during rush hours $t_{ij}^{busy}$ and free-flow travel time during midnight $t_{ij}^{free}$ to quantify traffic congestion
\begin{equation} 
    congestion_{ij}=\frac{t_{ij}^{busy}}{t_{ij}^{free}},
\end{equation}
whose value is larger or equal to one. 
Motivated by population-weighted transportation efficiency proposed in Ref. \cite{dong2016population}, we analyze the relation between biking-traffic-weighted DR and congestion, and discover a robust sublinear scaling relation
\begin{equation} \label{eq.scaling}
    \frac{T_{ij}}{T}\times\frac{t_{ij}^{busy}}{t_{ij}^{free}}=\left(\frac{T_{ij}}{T}\times DR_{ij}\right)^\alpha,
\end{equation}
where $T_{ij}$ is the dockless biking traffic on the route between locations, $T=\sum_{i,j}T_{ij}$ is the total biking traffic in the whole city, and the scaling exponent $\alpha$ is equal to 0.79, 0.79, and 0.83 for Beijing, Shanghai, and Xiamen, respectively (see Fig. \ref{Figure 5}(a)-(c)). 
A larger $\alpha$ indicates a stronger influence of DR and biking traffic over congestion levels. 
Among three cities, Xiamen has the largest $\alpha$ and this is in line with the fact that Xiamen is quite small and the road networks are sparser, which generally results in fewer routing choices and more concentrated traffic.  
Such a simple scaling relation reflects an intrinsic entanglement of travel modes (bike and vehicle) and topology of infrastructure networks, and this allows us to infer the traffic congestion on a specific route with just the biking traffic and DR
\begin{equation} \label{eq.traffic}
    \frac{t_{ij}^{busy}}{t_{ij}^{free}}=\left(\frac{T_{ij}}{T}\right)^{\alpha-1}\times DR_{ij}^\alpha.
\end{equation}
As $\alpha<1$, this indicates that less biking traffic and more detoured routes generally lead to more severe congestion. 
The scaling relation is nontrivial, as when we take off the biking traffic, the relation between congestion and DR is almost irrelevant (see Fig. \ref{Figure 5}(d)-(f)).

As the traffic congestion is related to the demand over supply \cite{ccolak2016understanding,xu2017clearer,xu2019unraveling} (i.e., volume over capacity), which is generally formulated as the Bureau of Public Roads (BPR) function
\begin{equation} \label{eq.BPR}
    t_{ij} = \left(1+\eta \left( \frac{V_{ij}}{C_{ij}} \right) \right)^\beta \times t_{ij}^{free}, 
\end{equation}
where $t_{ij}$ is the travel time by car given the vehicle traffic $V_{ij}$ and road capacity $C_{ij}$, $\eta$ and $\beta$ are coefficients that can be better estimated from empirical data, and $\beta$ is generally larger than one. Here, $C_{ij}$ can be estimated from metadata related to roads (e.g., the number of lanes, speed limit, road category) \cite{website:OSM}, and combining Eq. \ref{eq.traffic} and Eq. \ref{eq.BPR}, we can reveal the interplay between biking traffic and vehicle traffic and structure of road networks 
\begin{equation} \label{eq.VijTij}
     \frac{t_{ij}^{busy}}{t_{ij}^{free}} = \left(1+\eta \left( \frac{V_{ij}}{C_{ij}} \right) \right)^\beta = \left(\frac{T_{ij}}{T}\right)^{\alpha-1}\times DR_{ij}^\alpha. 
\end{equation}
Once the full sample vehicle traffic data are available, such a relation can be better tested. 
In practice, we assume that biking traffic and vehicle traffic may have a stronger influence on each other especially when there is no dedicated cycling lane. Biking traffic may highly probably slow down the vehicle traffic and the vehicle traffic may jeopardize the safety of cyclists, besides, exhaust emissions by cars would also harm the health of riders. 
In addition, the impacts of the structure of road networks are also entangled, for example, for very detoured and low-capacity routes, cycling might be indeed a better choice than driving \cite{shaheen2013public}. 


\begin{figure}[htbp] \centering
\subfigure[Beijing]{
\begin{minipage}[t]{.33\linewidth}\centering
\includegraphics[width=\linewidth]{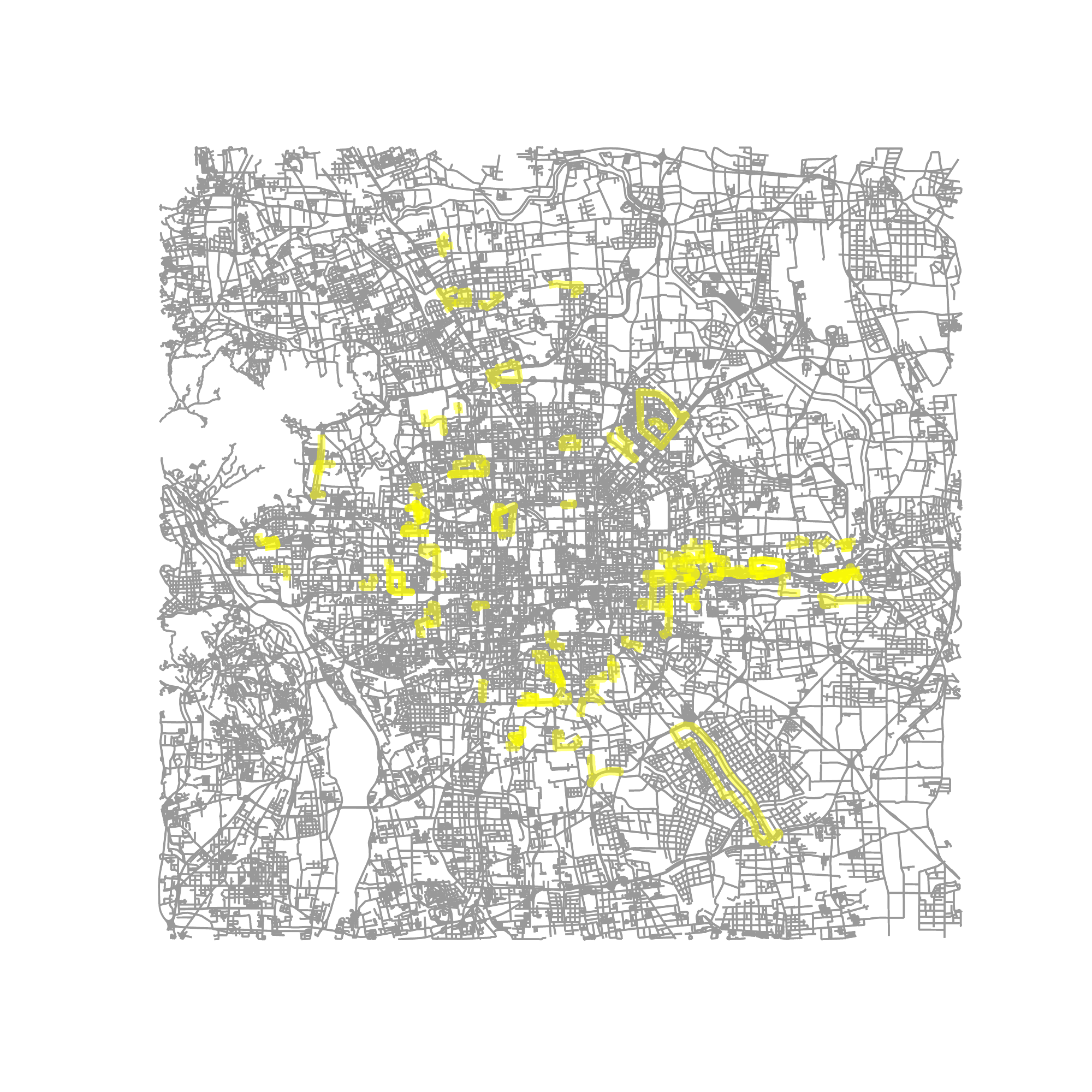}
\end{minipage}}%
\subfigure[Shanghai]{
\begin{minipage}[t]{.33\linewidth}\centering
\includegraphics[width=\linewidth]{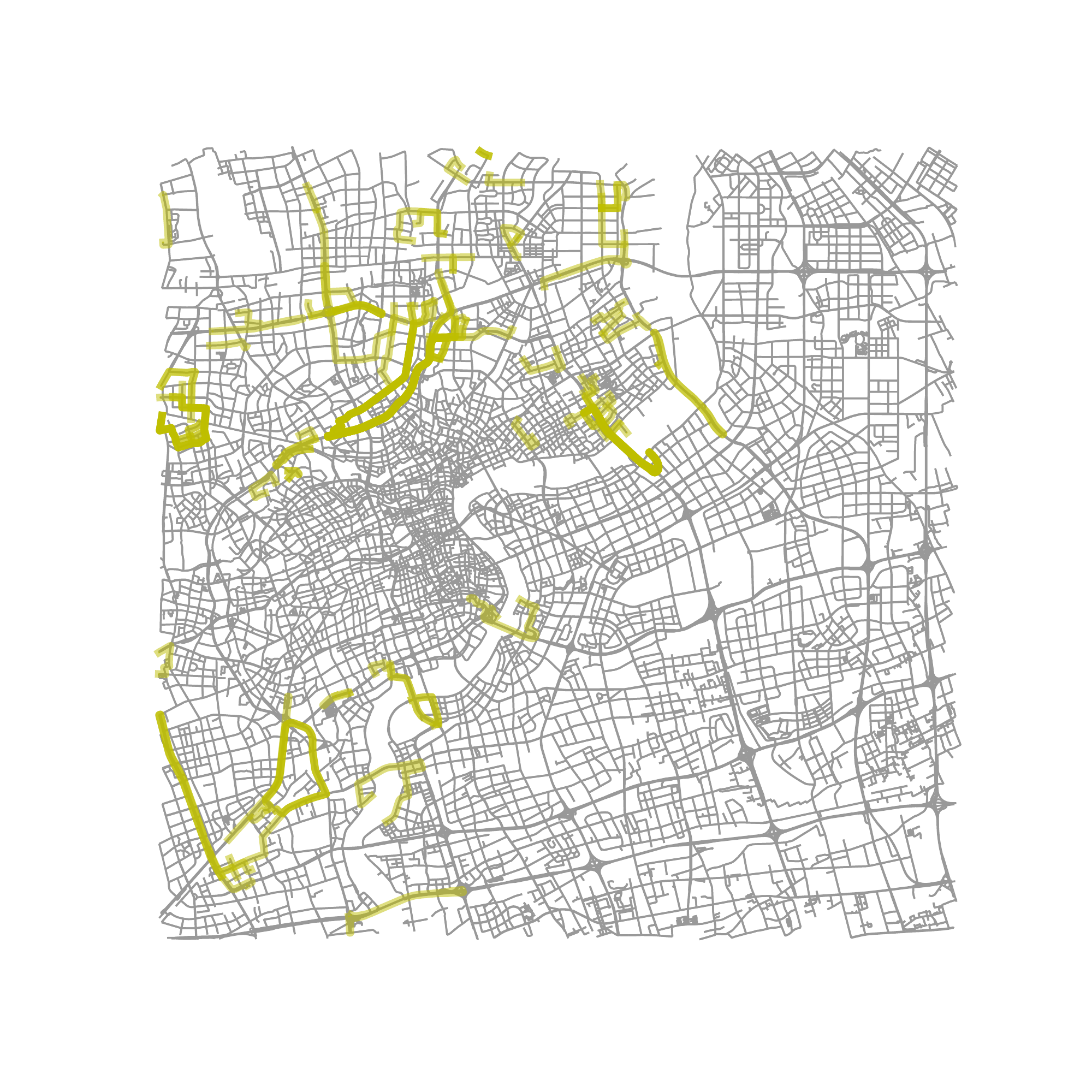}
\end{minipage}}%
\subfigure[Xiamen]{
\begin{minipage}[t]{.33\linewidth}\centering
\includegraphics[width=\linewidth]{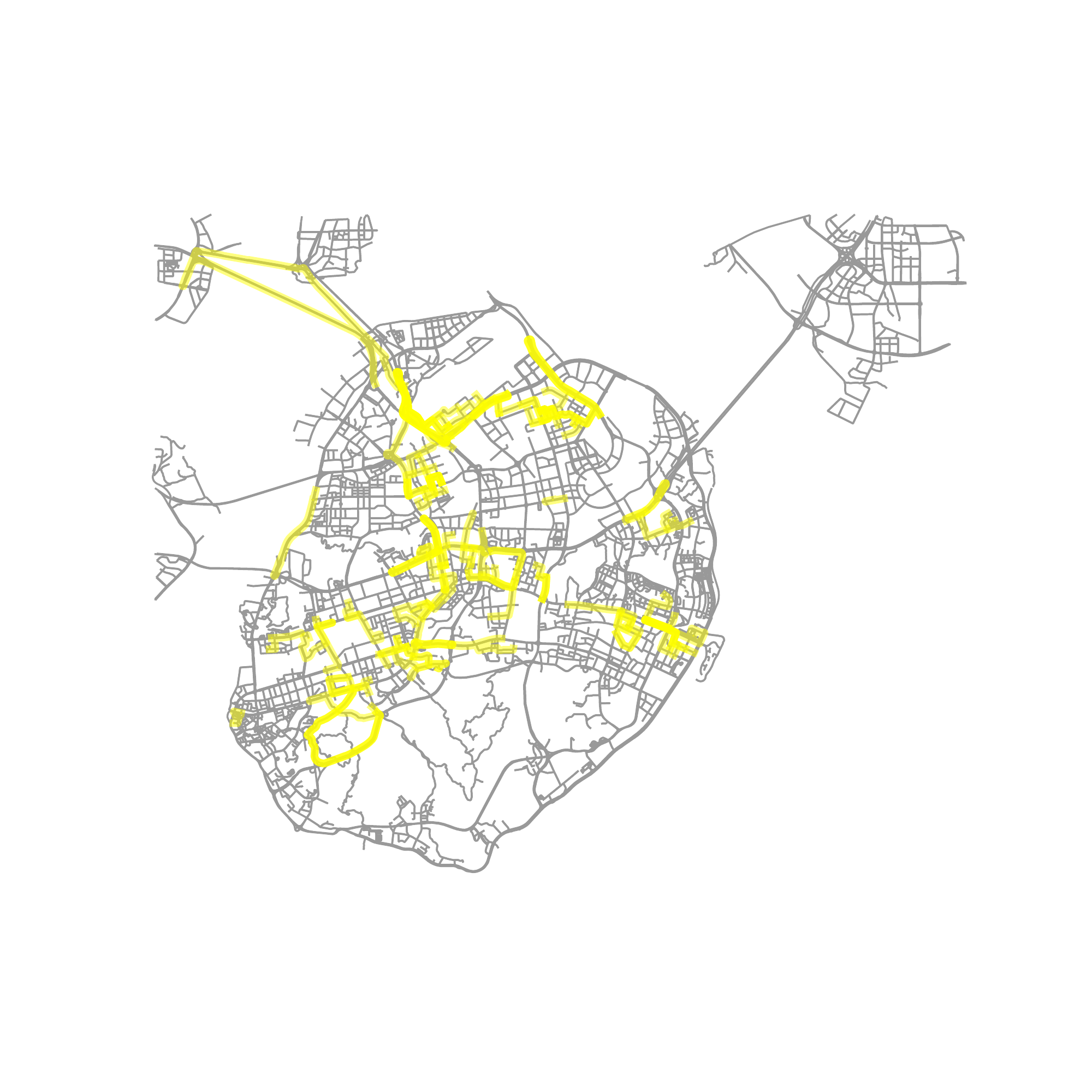}
\end{minipage}}%

\caption{Cycling routes that require prior improvements in three cities}
\label{fig.inefficientRoutes}
\end{figure}


\subsection{Identification of inefficient routes of biking}
Apart from traffic analysis, the traffic-weighted-DR (i.e., $\frac{T_{ij}}{T}\times\frac{r_{ij}}{d_{ij}}$) can be applied to detect inefficient biking routes. When a route is more detoured, it is generally not friendly to cycling, as it would be harder to navigate \cite{coutrot2022entropy}, and when many people are using such routes to commute or connect commuting, then it should be improved with a higher priority. Here, via OSMnx \cite{boeing2017osmnx}, we visualize the first one hundred inefficient cycling routes with the largest traffic-weighted-DR in each city (see Fig. \ref{fig.inefficientRoutes}(a)-(c)). 
Many inefficient routes are located at the fringe of cities, where infrastructure and public transportation might be relatively underdeveloped. Complex geographic or hydrological factors may also lead to inefficient routes, for example, the ones on the opposite sides of the Huangpu River in Shanghai or separated by Yundang Lake and Wangtai Mountain in Xiamen. Apart from segmenting effects posed by natural factors, man-made infrastructures might also pose similar effects. Wide roads might lead to ``n''-shaped long detours when the origin and destination locations are on the opposite side (e.g., near the CBD of Beijing that location around East 3rd Ring Road to East 4th Ring Road,  where travel demands posed by working population there are also high). 
Some large street blocks (e.g., large gated working units or residential communities, or even airports) also leads to encircled long detours. Compared to opening up gated regions for vehicle traffic, it might be viable to establish a biking-traffic-only lane through them (e.g., working units, residential communities \cite{li2021heightened}, and universities). 

\section{Conclusion and Discussion}
In this work, we reveal a hidden scaling relation between biking-traffic-weighted detour ratio and congestion, which signifies entangled relation between travel modes (e.g., biking and driving) and structure of road networks (e.g., detour ratio, road capacity) and the heterogeneous spatio-temporal travel demands. Such a simple scaling relation helps us on gaining a better understanding of complex urban systems and can be aggregated into more sophisticated models for congestion prediction with limited data. In addition, the biking-traffic-weighted detour ratio also can be used to detect inefficient routes for cycling, which are mainly caused by enclosed large regions, insufficient development of the infrastructure network, and segmentation effects posed by wide roads. 
Our findings would help alleviate urban congestion, make better urban planning, and improve transportation efficiency in cities.

Possible further research includes the following aspects. Currently, most state-of-the-art collective mobility models \cite{simini2012universal,yan2014universal,yan2017universal} neglect the impacts of travel modes. Better understanding the generating mechanics behind biking traffic and devising collective mobility models suitable for cycling worth closer investigations, as they will be crucial for predicting the evolution of the system after improving certain inefficient routes. 
In addition, although bikes have been the most affordable transportation conveyance, the influence of income on cycling activities via the dockless sharing bike still can be obvious at the collective level \cite{gao2022quantifying}, which worth future investigations for its impacts on transportation systems. There is a common concern that the expansion of fast transportation networks might bring about gentrification and further chase away the poor \cite{kahn2007gentrification}. A better managed cycling infrastructure network \cite{olmos2020data,szell2022growing} would be an important component towards a more sustainable and equal transportation system \cite{liu2021exploring}. 




\section*{Acknowledgements}
We acknowledge financial support from the National Natural Science Foundation of China (Grant No. 61903020), Fundamental Research Funds for the Central Universities (Grant No. buctrc201825).

\appendix
\renewcommand\thefigure{A.\arabic{figure}}    

\begin{figure}[!htbp] \centering
\subfigure[Beijing]
{    \begin{minipage}[b]{.3\linewidth}        \centering
        \includegraphics[scale=0.25]{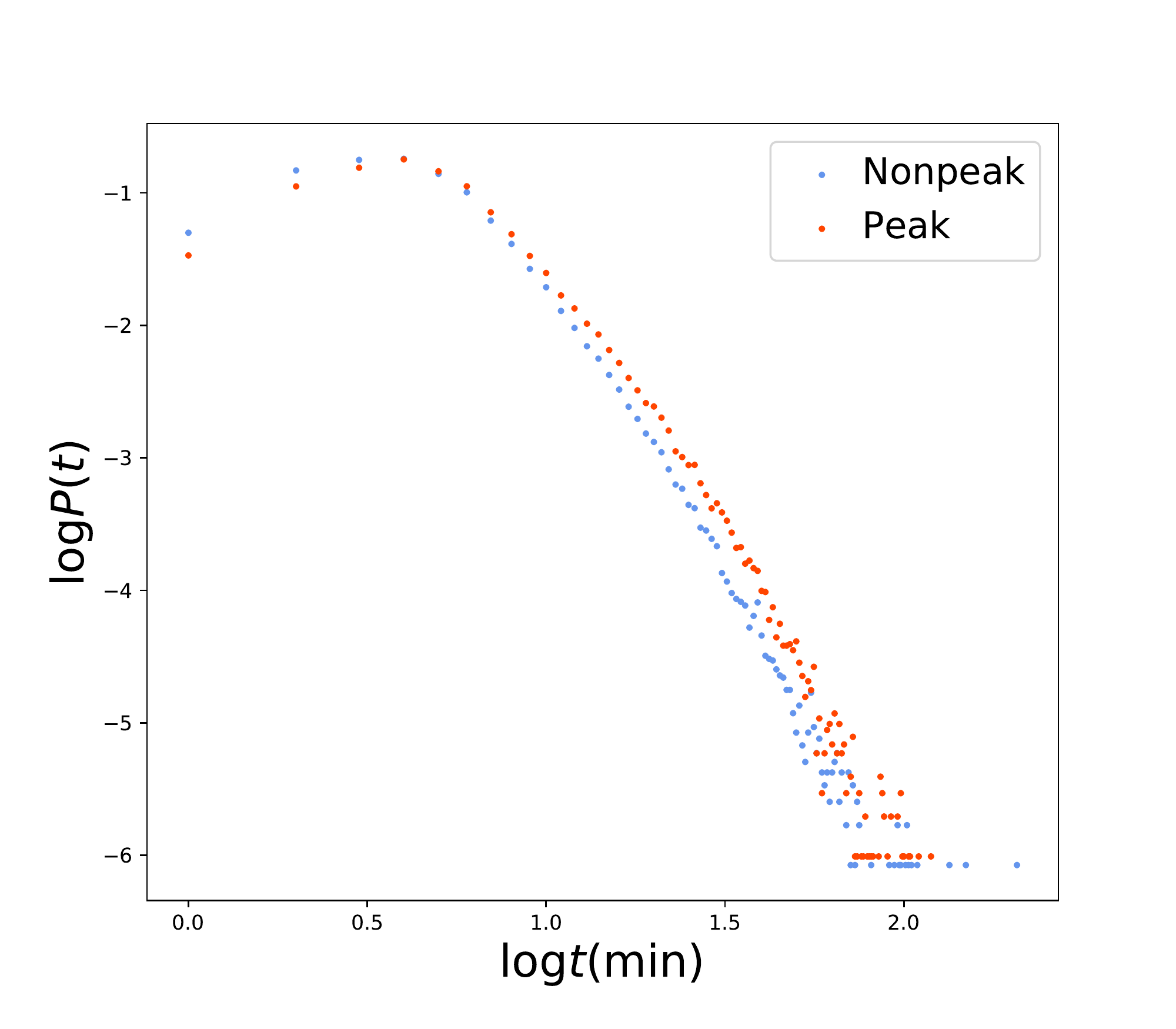}
    \end{minipage} }
\subfigure[Shanghai]
{ 	\begin{minipage}[b]{.3\linewidth}        \centering
        \includegraphics[scale=0.25]{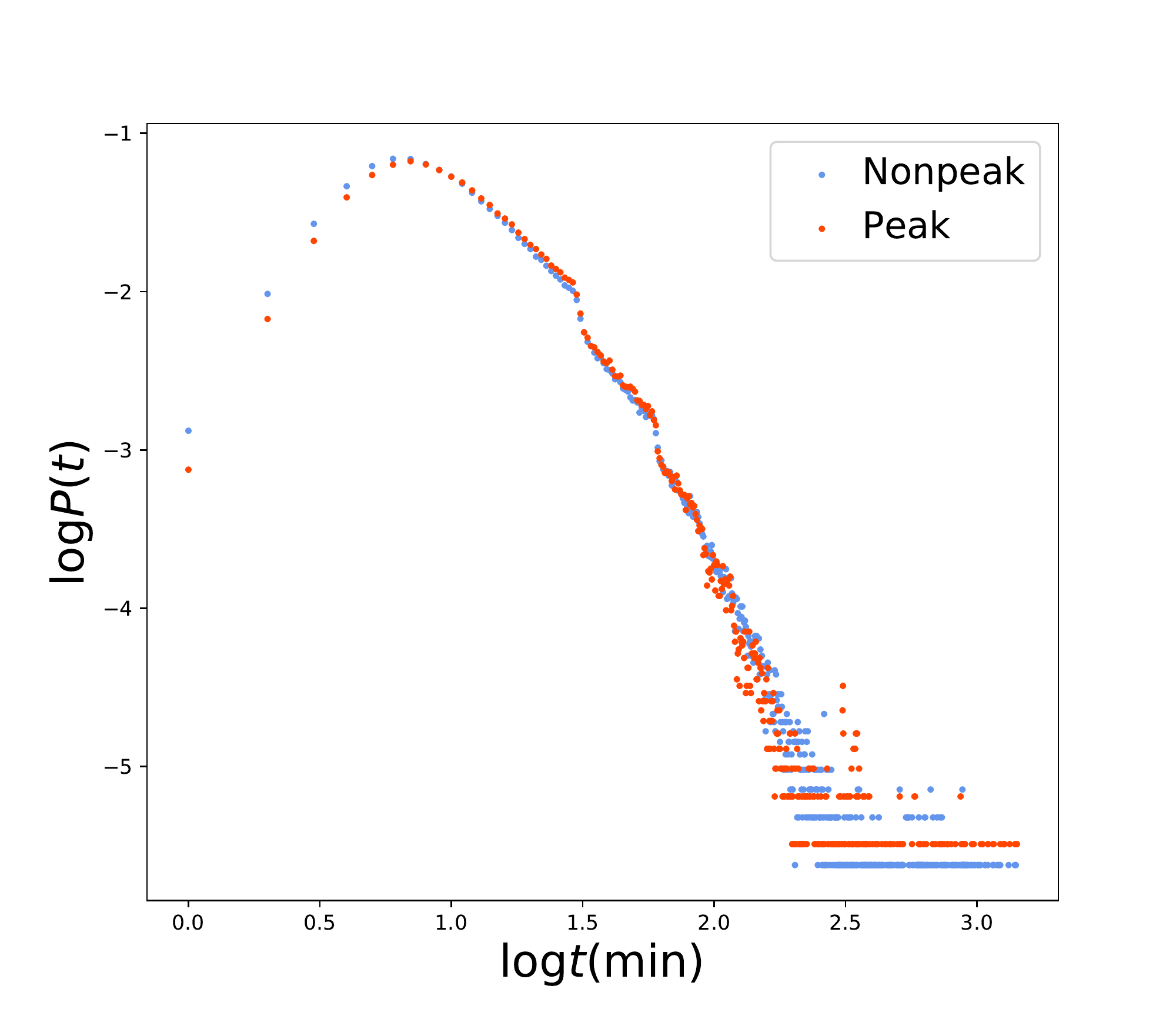}
    \end{minipage}}
\subfigure[Xiamen]
{ 	\begin{minipage}[b]{.3\linewidth}        \centering
        \includegraphics[scale=0.25]{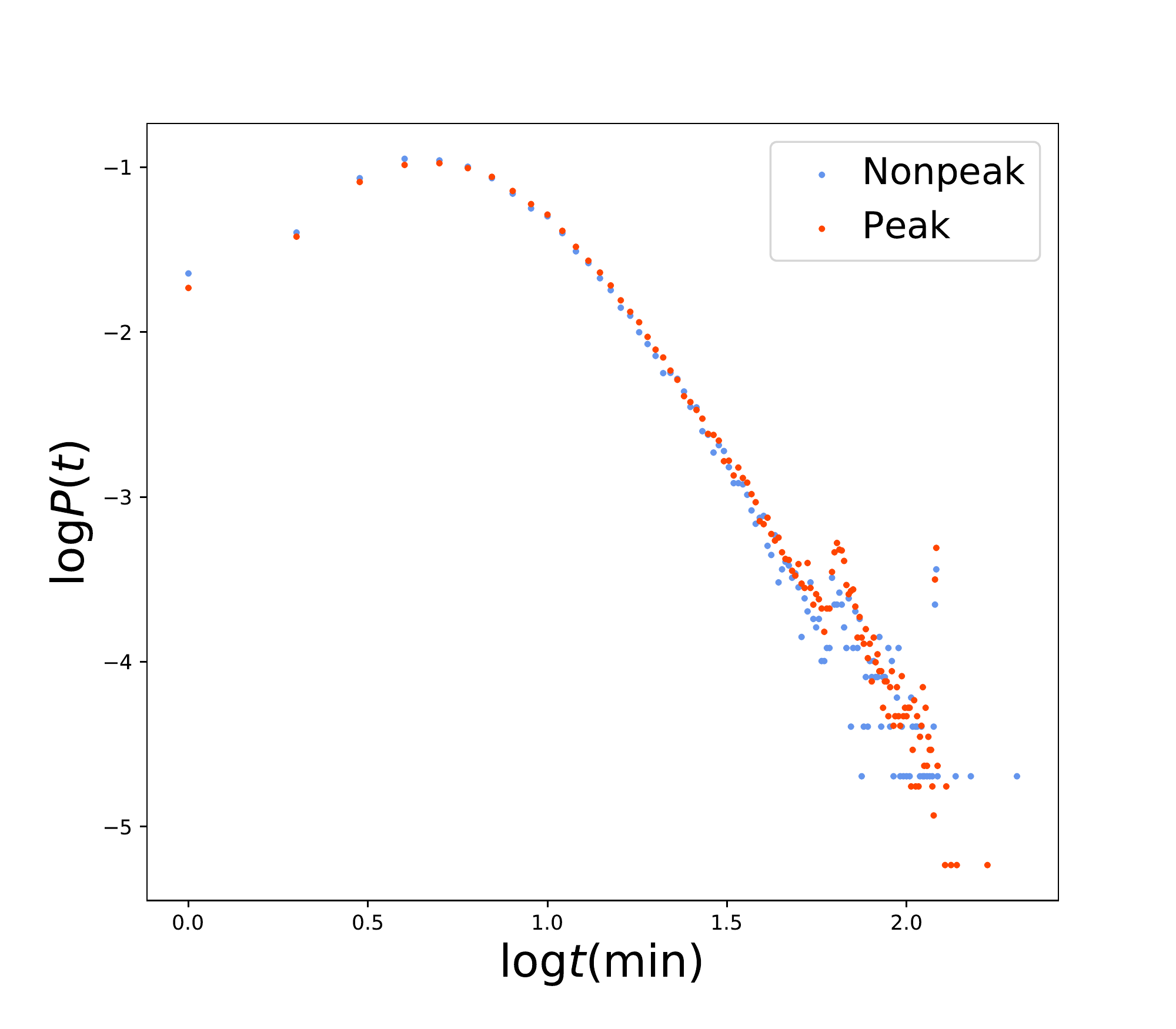}
    \end{minipage}}
\caption{Distribution of cycling duration in three cities.}
\label{Figure 2}
\end{figure}

\section*{Appendix A. Cycling duration distribution}
The distributions of cycling duration peak roughly between 3-6 minutes, but also exhibit power-law fat tails \cite{brockmann2006scaling} across cities (see Fig. \ref{Figure 2}). 
This suggests that the average riding duration is not as representative as ordinarily assumed \cite{chen2019analyzing}, and there is a notable fraction of trips of a much longer riding duration than the average when compared to a Poisson or Normal distribution with the same mean value. 
Given the fact that during rush hours, there are more longer distance trips. Collapsing of cycling duration distributions for rush and off-peak hours in Shanghai indicates that cyclists might ride relatively faster during the rush hours (see Fig. \ref{Figure 1}(b) and Fig. \ref{Figure 2}(b)). 
Note that the Beijing dataset dose not provide the actual arrival time of each trip, we make estimations based on expected cycling duration queried from Amap API (\url{https://lbs.amap.com/}) for the shortest route between given origin and destination locations, which will be trivially proportional to the routing distance. 
Interestingly, the traveling characteristics of cycling activities are just on the contrary to those taking taxis. The taxi tends to have a shorter trip distance and duration during rush hours than during off-peak hours \cite{feng2022scaling}. 

\renewcommand\thefigure{B.\arabic{figure}}    

\section*{Appendix B. Road width distribution}
\begin{figure}[!htbp]     \centering
    \includegraphics[scale=0.37]{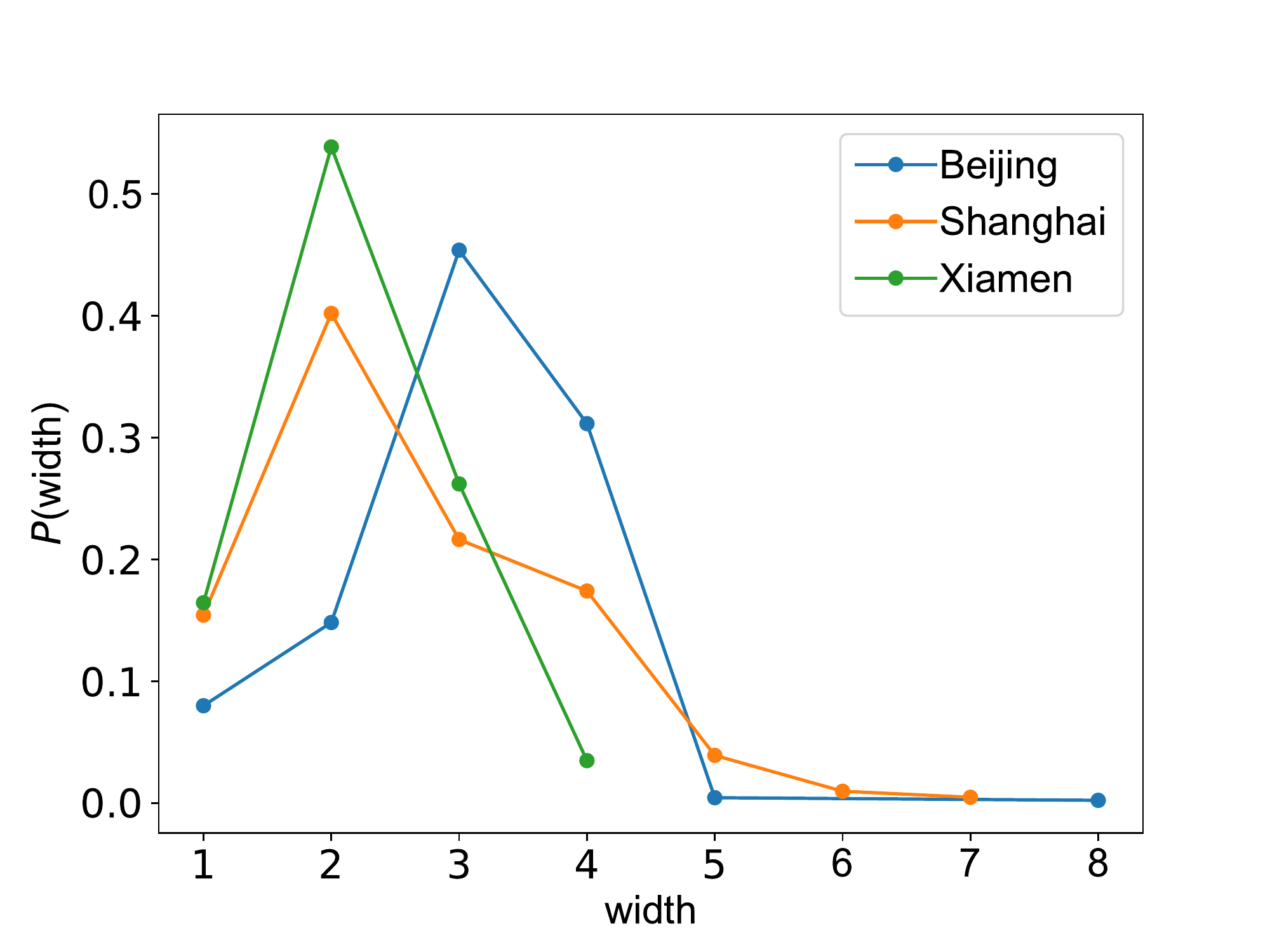}
    \caption{Distribution of road width (the number of lanes) in three cities. The road network data is obtained from OpenStreetMap \cite{website:OSM}}
    \label{fig:Road width}
\end{figure}

We use the number of vehicle lanes of a road to approximate its width (see Fig. \ref{fig:Road width}). The road networks  for Beijing, Shanghai, and Xiamen are obtained from OpenStreetMap \cite{website:OSM} via the OSMnx \cite{boeing2017osmnx} in Python. Each road is associated with a variety of metadata, including the category of the road, the speed limit, and the number of lanes. Here, we consider roads within 15 km of the city center, which roughly covers the region of cycling activities. 

\end{document}